\documentclass[prd,aps,10pt,twocolumn,showpacs,reprint,nofootinbib]{revtex4}
 
\usepackage{graphicx}
\usepackage{amssymb}
\usepackage{filecontents}
\usepackage{amsmath}
\usepackage{color}
\usepackage{float}
\RequirePackage[colorlinks,citecolor=blue,urlcolor=blue,linkcolor=blue]{hyperref}
\begin{document}
\title{Generalized Energy-Momentum-Squared Gravity in the Palatini Formalism}
\author{Elham Nazari$^{1}$}
\author{Farahnaz Sarvi$^{1}$}
\author{Mahmood Roshan$^{1,2}$\footnote{mroshan@um.ac.ir}}

\affiliation{$^1$Department of Physics, Ferdowsi University of Mashhad, P.O. Box 1436, Mashhad, Iran\\
$^2$ School of Astronomy, Institute for Research in Fundamental Sciences (IPM), 19395-5531, Tehran, Iran }

\begin{abstract}
We study the generalized version of energy-momentum squared gravity (EMSG) in the Palatini formalism. This theory allows the existence of a scalar constructed with energy-momentum tensor as $T_{\alpha\beta}T^{\alpha\beta}$ in the generic action of the theory. We study the most general form of this theory in the Palatini framework and present the underlying field equations. The equations of motion of a massive test particle have been derived. The weak field limit of the theory is explored and the generalized version of the Poisson equation is obtained. Moreover, we explore the cosmological behavior of the theory with emphasis on bouncing solutions.  Some new bouncing solutions for the specific Palatini EMSG model given by the Lagrangian density $\mathcal{L}=R+\beta R^2+\eta T_{\mu\nu}T^{\mu\nu}$ are introduced. We show that only the case $\eta>0$ can lead to viable cosmic bounce.
\end{abstract}
\maketitle
\section{Introduction}
Recently the energy-momentum squared gravity (EMSG)\footnote{EMSG is used only for the special case with Lagrangian density $\mathcal{L}=R+\eta T_{\mu\nu}T^{\mu\nu}$. However, for convenience throughout this paper, we use this name also for the generalized version of the theory i.e., $\mathcal{L}=f(R,T_{\mu\nu}T^{\mu\nu})$. } has been introduced to resolve the Big Bang singularity \cite{roshan2016energy,Arik:2013sti}. This theory allows the existence of a special scalar constructed using the energy-momentum tensor as $T_{\alpha\beta}T^{\alpha\beta}$. This is different from theories exploiting the trace of $T_{\mu\nu}$, for example, see \cite{Harko:2011kv}. The main idea behind EMSG was to avoid the existence of the Big Bang singularity. It is shown in \cite{roshan2016energy} that a simple model of EMSG in the metric formulation leads to a viable cosmic behavior possessing a true sequence of the cosmic epochs. On the other hand, the only difference with the standard picture is that there is no early universe singularity. In other words, EMSG predicts a minimum length and a maximum but finite energy density in the early universe. Currently, interest in this theory is not limited to removing the Big Bang singularity. Depending on the EMSG model considered, it may suggest interesting modifications/corrections to the whole cosmic history and not only to the early universe. Although this model does not avoid black hole singularities, it predicts larger masses for neutron stars with the ordinary equation of states \cite{Nari:2018aqs,Akarsu:2018zxl}. For a recent work on quark compact stars in EMSG we refer the reader to \cite{Singh:2020bdv}. 

Recently, several aspects of EMSG have been investigated. In \cite{Bahamonde:2019urw}, different models of EMSG have been studied using the dynamical system approach. Consequently, by analyzing the relevant fixed points, interesting cosmological behavior has been reported. For other investigations on the cosmology of EMSG, we refer the reader to \cite{Board:2017ign}-\cite{Akarsu:2019ygx}. In \cite{kazemi2020jeans}, the Jeans analysis has been explored in EMSG and a new Jeans mass has been introduced. Furthermore, by finding a generalized version of Toomre's parameter in the context of EMSG, the local stability of hyper-massive neutron stars has been studied and some constraints on the free parameters of the theory have been reported. Recently it has been shown in \cite{Barbar:2019rfn} that the bounce reported in \cite{roshan2016energy} cannot describe the real Universe in the sense that it cannot regularly connect the early universe bounce to a viable de Sitter late time universe. More specifically EMSG has a viable bounce only in the matter dominated cosmological toy model \cite{Barbar:2019rfn}. To achieve a viable bounce, the specific term $(T_{\mu\nu}T^{\mu\nu})^{5/8}$has been included in the action instead of the quadratic scalar $T_{\mu\nu}T^{\mu\nu}$. However, it is straightforward to show that the existence of a vacuum energy density in EMSG, can resolve the mentioned problem without any changes to the action of the theory\footnote{Private communication with Fatimah Shojai.}. It is necessary to mention that EMSG discriminates between geometrical cosmological constant appearing in the action and the vacuum energy density treated as a perfect fluid. Albeit, in this case although there is a regular bounce connecting the early universe bounce to the late time de Sitter phase, the viability of the cosmological model needs more careful investigations. 

So far, all the research on EMSG has been carried in the metric formulation. On the other hand, like any other modified theory of gravity, the Palatini framework, in principle, leads to a different theory with different predictions. Therefore, our main purpose in this paper is to explore EMSG in the Palatini formulation. In this formulation, the connection $\Gamma_{\mu\nu}^{\alpha}$ and the metric tensor $g_{\mu\nu}$ are considered as different and independent fields. As we know, Palatini formulation of general relativity (GR) coincides with the metric one \cite{d1992introducing}-\cite{Olmo:2011uz}. However, this is not the case necessarily for other gravitational theories, for example in the case of $f(R)$ gravity see \cite{Sotiriou:2008rp}.

We explore the weak field limit by finding the generalized form of the Poisson equation. Furthermore, we show that Palatini EMSG is also capable to introduce cosmological bouncing solutions. All the necessary conditions for having a bounce are found.

The content of the paper is organized as follows.  We begin in Sec. \ref{sec:eom} by obtaining the modified version of the Einstein field equations in the Palatini formalism of EMSG. Then, by using these field equations, we find the conservation and geodesic equations in this framework. In Sec. \ref{sec:wf}, we investigate the weak-field limit of this model. Also, in Sec. \ref{Cosmicbounce}, by choosing a toy model given by the Lagrangian density $\mathcal{L}=R+\beta R^2+\eta T_{\mu\nu}T^{\mu\nu}$, we study the existence of a cosmic bounce in the very early universe. Finally, in Sec. \ref{sec:con}, our results are discussed.

Throughout this paper, the Greek indices vary from 0 to 3. Also, the Latin indices vary from 1 to 3. Here, we assume that the metric signature is $(-,+,+,+) $ and we use physical units in which the velocity of light is $c=1$.
\section{Field equations in Palatini EMSG } 
\label{sec:eom}
We launch our calculation of the field equations in the torsionless Palatini formulation of EMSG by introducing the following action that is a function of a metric as well as an independent connection. 

\begin{equation}\label{eq:action}
\mathcal{S}=\frac{1}{2\kappa} \int d^4x \sqrt{-g} f(R,Q) + \int d^4x \sqrt{-g} \mathcal{L}_m(g_{\mu\nu},\psi), 
\end{equation}
where $\kappa = 8 \pi G$ and we work using the units in which the velocity of light is $c=1$. Here, $g$ is the determinant of the spacetime metric tensor $g_{\mu\nu}$, and $\psi $ is the representative of the matter fields. In the EMSG, we assume that $f$ is an arbitrary function of $Q = T_{\mu\nu}T^{\mu\nu}$ and the Ricci scalar $R = g^{\mu\nu}R_{\mu\nu}$. 
It should be mentioned that in the Palatini formalism where the spacetime metric and affine connection $\Gamma^\lambda_{\mu\nu}$ are considered as independent fields, the Riemann and Ricci tensors are constructed from the affine connection $\Gamma^\lambda_{\mu\nu}$.
So, we have $R_{\mu\nu}(\Gamma) = {R^\alpha}_{\mu\alpha\nu}(\Gamma)$ in which $g_{\mu\nu}$ is not utilized to raise or lower indices.
Here, as usual, we assume that the matter Lagrangian density $\mathcal{L}_m$ is only a function of $\psi$ and the metric components and it does not depend on the metric derivations. Although the matter field is in principle coupled to the metric, we assume that there is no coupling between the matter fields $\psi$ and the connection $\Gamma^\lambda_{\mu\nu}$.

By considering the above assumptions and variating the action (\ref{eq:action}) with respect to the metric and affine connection, after some simplification, we arrive at
\begin{equation} \label{pvariation}
\begin{split}
&\delta \mathcal{S}=\frac{1}{2\kappa} \int d^4x \sqrt{-g} f_{R}  g^{\mu\nu} \delta R_{\mu\nu}(\Gamma)\\
+&\frac{1}{2\kappa} \int d^4x \sqrt{-g}\Big( f_{R} R_{\mu\nu}+f_{Q} \Theta_{\mu\nu}-\frac{1}{2}f g_{\mu\nu}  -\kappa T_{\mu\nu} \Big)\delta g^{\mu\nu}, 
\end{split}
\end{equation}
where $f_R = \partial f/ \partial R$ and $f_{Q} = \partial f/\partial Q$. Here, $T_{\mu\nu}$ and $\Theta_{\mu\nu}$ are defined by
\begin{equation}\label{tmu}
T_{\mu\nu}=-\frac{2}{\sqrt{-g}} \frac{\delta (\sqrt{-g} \mathcal{L}_m) }{\delta g^{\mu\nu}}, ~~~~\Theta_{\mu\nu} \equiv \frac{\delta Q}{\delta g^{\mu\nu}}.
\end{equation}

The modified version of the field equations is then derived from the condition that $\delta \mathcal{S} =0$. Therefore, the first integral in Eq. \eqref{pvariation} representing the variation of action with respect to the affine connection can be reduced to the following equation
\begin{equation} \label{eq:connection}
\nabla_{\lambda}^{\Gamma}(\sqrt{-g} f_R g^{\mu\nu})=0,
\end{equation}
after using the identity 
$\delta R_{\sigma\nu}=\nabla_\rho\big(\delta \Gamma^\rho_{\nu\sigma}\big)-\nabla_\nu\big(\delta \Gamma^\rho_{\rho\sigma}\big)$
and applying the divergence theorem.
Here, $\nabla_{\lambda}^{\Gamma}$ stands for the covariant derivative which is associated with the independent symmetric connection $\Gamma^\lambda_{\mu\nu}$. We recall that there is no torsion postulated in this theory. The general version of Eq. \eqref{eq:connection} containing both symmetric and antisymmetric parts of $\Gamma^\lambda_{\mu\nu}$ is derived in \cite{Afonso:2017bxr}. 
The second sector of Eq. \eqref{pvariation} denoting the variation of action with respect to the metric can be written as
\begin{equation}\label{eq:metric}
f_R R_{\mu\nu} - \frac{1}{2}f g_{\mu\nu}=\kappa \tau_{\mu\nu},
\end{equation}
in which $\tau_{\mu\nu}$ is an \textit{effective} energy-momentum tensor defined as
\begin{equation} \label{eq:tau}
\tau_{\mu\nu}=T_{\mu\nu}- \frac{f_{Q}}{\kappa} \Theta_{\mu\nu}.
\end{equation}
Eq. \eqref{eq:metric} is the modified version of the Einstein field equations in the Palatini formulation of EMSG.

Now, by utilizing the similar method introduced in \cite{Barrientos:2018cnx}, we attempt to simplify Eqs. \eqref{eq:connection} and \eqref{eq:metric}.
To do so, we first contract Eq. \eqref{eq:metric} with the metric tensor $g^{\mu\nu}$ and obtain the following trace equation
\begin{equation} \label{eq:traceEq}
R f_R  - 2f=\kappa \tau,
\end{equation}
in which $\tau = g^{\mu\nu} \tau_{\mu\nu}$. We can then introduce the conformal relation between the physical metric $g_{\mu\nu}$ and auxiliary metric\footnote{Although we call $h_{\mu\nu}$ as auxiliary metric throughout this manuscript, it should be noted that, as seen from Eq. \eqref{gdelta}, the perturbations of the physical metric $g_{\mu\nu}$ couple to those of $h_{\mu\nu}$. Therefore, studying the gravitational wave and investigating its propagation in the linearized theory of Palatini EMSG would reveal whether the auxiliary metric $h_{\mu\nu}$ can be ``physical" or not. For instance, in \cite{Chen:2018vuw}, the gravitational waves in the Palatini-type gravity theories are studied and from this point of view, the physical aspect of the metric $h_{\mu\nu}$ is investigated. } $h_{\mu\nu}$ as $h_{\mu\nu}=f_R g_{\mu\nu}$ 
for which one can show that 
\begin{equation}\label{gamma}
\Gamma^\lambda_{\mu\nu}= \frac{h^{\lambda\alpha}}{2}\left(\partial_\mu h_{\alpha\nu}+\partial_\nu h_{\alpha\mu}-\partial_\alpha h_{\mu\nu}\right).
\end{equation}
For more detail, see \cite{Barrientos:2018cnx} and references therein.
The above relation implies that the affine connection $\Gamma^\lambda_{\mu\nu}$ coincides with the Christoffel symbols of the auxiliary metric $h_{\mu\nu}$.
Next, by applying the definition of $h_{\mu\nu}$ and after some manipulation, we can rewrite the connection equations \eqref{eq:connection} and field equations \eqref{eq:metric} as
\begin{equation}\label{h}
\nabla_{\lambda}(\sqrt{-h} h^{\mu\nu})=0,
\end{equation}
and 
\begin{equation} \label{eq:Riccieqs}
{R^\mu}_{\nu}(h)=\frac{\kappa}{f_R^2} \Big({\tau^\mu}_{\nu} + \frac{f}{2\kappa} {\delta^\mu}_{\nu}\Big), 
\end{equation}
respectively. Here,  ${R^\mu}_{\nu}(h) = h^{\mu\alpha}R_{\alpha \nu}$.

Although the appearance of the latter equation is similar to the corresponding relation in the metric-affine $f(R, T)$ theories \cite{Barrientos:2018cnx}, where $T$ is the trace of the energy-momentum tensor, it should be recalled since the role of $Q$ is encoded in $\tau_{\mu\nu}$ and $f$, the equation of motion is different. Eq. \eqref{eq:Riccieqs} is our final field equation for Palatini EMSG and in the following sections we will explore its consequences in different contexts. To do so, we use the perfect fluid as the matter source.  In this case, the energy-momentum tensor is given by
\begin{equation} \label{eq:pftumunu}
T_{\mu\nu}=(\rho + p) u_{\mu}u_{\nu}+p g_{\mu\nu}, 
\end{equation}
where $u^{\mu}$ is the velocity four vector for which $u_{\mu}u^{\mu}=-1$. Here, $\rho$ and $p$ are the energy density and pressure of the fluid, respectively. It is necessary to mention that except the fifth force and weak field limit of the geodesic equation where we study two different matter Lagrangian densities, i.e., $\mathcal{L}_m=p$ and $\mathcal{L}_m=-\rho$, we assume everywhere throughout this paper that $\mathcal{L}=p$. In this case, for the perfect fluid, it is straightforward to verify that 
\begin{equation}
\Theta_{\mu\nu}=-(\rho^2+4 \rho p+3 p^2) u_{\mu}u_{\nu},
\end{equation}
Substituting these relation into (\ref{eq:tau}), we simplify Eq. \eqref{eq:Riccieqs} as follows
\begin{equation} \label{eq:RiccieqsFluid}
\begin{split}
{R^\mu}_{\nu}(h)=\frac{\kappa}{f_{R}^2} \bigg[\big(\rho&+p\big) u^{\mu} u_{\nu}+\Big(p+\frac{f}{2 \kappa}\Big)
 \delta^{\mu}_{\nu}\\
& +\frac{f_{Q}}{\kappa} \Big(\rho^2+4 \rho p+3 p^2\Big)u^{\mu} u_{\nu}\bigg].
\end{split}
\end{equation}

As a final task in this part, let us find the modified version of the Einstein curvature tensor in the Palatini EMSG. To do so, by utilizing Eq. \eqref{eq:Riccieqs}, we obtain the Ricci scalar. Then by considering the definition of the Einstein tensor and some algebra, we have
\begin{equation}\label{eq:Einstein}
{G^\mu}_{\nu}(h)=\frac{\kappa}{f_R^2}\bigg[{\tau^\mu}_{\nu}-\frac{\delta^{\mu}_{\nu}}{2} \Big(\tau + \frac{f}{\kappa}\Big)\bigg].
\end{equation}
By specifying the function $f(R, Q)$ and the energy-momentum  tensor of the matter fields, in principle, one may use these field equations for obtaining the auxiliary metric $h_{\mu\nu}$. Consequently, regarding the conformal relation between $h_{\mu\nu}$ and $g_{\mu\nu}$ the physical spacetime metric in the Palatini EMSG can be found.
The field equations \eqref{h} and \eqref{eq:Einstein} are the cornerstone of future calculations in the metric-affine $f(R,Q)$ theories.  

It is necessary to mention that there is no a priori way to find the generic function $f(R, Q)$. Normally these kinds of functions in modified gravity theories, are postulated in such a way that the theory reveals a new feature in comparison with GR while leaving the classical tests of the gravity unchanged. This new feature would be used to alleviate GR's problems. Some times forcing the theory to comply with some symmetry issues is helpful to fix the functionality of the generic function. For example see \cite{Roshan:2008tt} in which by requiring the Noether symmetry in Palatini $f(R)$  gravity, it is possible to fix the function $f(R)$.

\subsection{Conservation equations of $T_{\mu\nu}$}

As we know, the energy and momentum are conserved when the covariant derivative of the energy-momentum tensor of the matter vanishes. It should be noted that the covariant derivative should be written in terms of the physical metric $g_{\mu\nu}$. Therefore, in order to find the conservation equations in this theory, we first take the covariant derivative $\nabla_{\mu}^{(h)}$ of Eq. \eqref{eq:Einstein} where $\nabla_{\mu}^{(h)}$ is based on the auxiliary metric $h_{\mu\nu}$. Then by considering the Bianchi identity and obtaining the relation between $\nabla_{\mu}^{(h)}$ and $\nabla_{\mu}^{(g)}$, we find the conservation equations in this theory.

Applying the Bianchi identities, i.e., $\nabla_{\mu}^{(h)}{G^\mu}_{\nu}(h)=0$, reveals that the right-hand side of Eq. \eqref{eq:Einstein} can be written as
\begin{equation}\label{eq:coneq1}
\begin{split}
&\nabla_{\mu}^{(h)}{\tau^\mu}_{\nu}-\partial_{\nu}\Big(\frac{\tau} {2} +\frac{f}{2\kappa}\Big)\\
&-2\partial_{\mu} \ln f_R \bigg[{\tau^\mu}_{\nu}-\frac{\delta^{\mu}_{\nu}}{2}\Big(\tau + \frac{f}{\kappa}\Big)\bigg]=0.
\end{split}
\end{equation}
To simplify this relation, as mentioned before, one should rewrite $\nabla_{\mu}^{(h)}$ with respect to $\nabla_{\mu}^{(g)}$. To do so, let us find the relation between the Christoffel symbols of the auxiliary metric, $\Gamma^\lambda_{\mu\nu}$, and those of the physical metric, $\hat{\Gamma}^\lambda_{\mu\nu}$. One can easily show that 
\begin{equation}\label{gp}
\Gamma^\lambda_{\mu\nu}=\hat{\Gamma}^\lambda_{\mu\nu}+C^{\lambda}_{\mu\nu},
\end{equation}
after inserting the conformal relation between $h_{\mu\nu}$ and $g_{\mu\nu}$ within Eq. \eqref{gamma}. Here, $C^{\lambda}_{\mu\nu}$ is given by 
\begin{equation}\label{22}
C_{\mu\nu}^{\lambda}=\frac{f_{RR}}{2 f_R}\left(\delta^{\lambda}_{\nu} \partial_{\mu}R+\delta^{\lambda}_{\mu}\partial_{\nu}R-g^{\lambda\alpha} g_{\mu\nu} \partial_{\alpha}R\right).
\end{equation}
Now, we can use the above relations to transform the covariant derivative based on the auxiliary metric to the standard form. So, we have  
\begin{equation} \label{16p}
\nabla_{\mu}^{(h)}{\tau^\mu}_{\nu}=\nabla_{\mu}^{(g)}{\tau^\mu}_{\nu}+C_{\mu\lambda}^{\mu}{\tau^\lambda}_{\nu}-C_{\mu\nu}^{\lambda}{\tau^\mu}_{\lambda}.
\end{equation}
By considering the definition \eqref{22} and substituting Eqs. \eqref{16p} and \eqref{eq:traceEq} into Eq. \eqref{eq:coneq1}, after some manipulation and simplification, we finally arrive at
\begin{equation} \label{eq:contau}
\nabla_{\mu}^{(g)}{\tau^\mu}_{\nu}= -\frac{f_{Q}}{2\kappa}\partial_\nu Q.
\end{equation}
The above relation represents the conservation equation in the Palatini formulation of EMSG. We see that $T_{\mu\nu}$ is not conserved in Palatini EMSG. This is also the case in the metric formulation of EMSG \cite{roshan2016energy}. On the other hand, the effective energy-momentum tensor can be conserved provided that $\partial_\nu Q=0$. This is not a physically interesting condition. Therefore, we conclude that neither $T_{\mu\nu}$  nor $\tau_{\mu\nu}$ are conserved in Palatini EMSG
\footnote{Albeit, in the following, we show that for the null dust fluid, $T_{\mu\nu}$ is conserved in Palatini EMSG.}.
By setting $f_{Q}=0$, the standard energy-momentum tensor conservation is recovered.

This non-conservation has immediate consequences that could rule out the theory. Therefore, this kind of corrections to the gravitational law should be carefully treated. In the next subsection, we explore this issue with more details.

\subsection{Fifth force } \label{Fifth force}
In this subsection, we obtain the geodesic equation in Palatini EMSG. In fact, we investigate the world line of a test particle and show that the path deviates from the standard geodesic curves of the spacetime. To be specific, we show that the right-hand side of the geodesic equation is nonzero and behaves like an extra force, namely the \textit{fifth force}.

To find this extra fore, we apply the method presented in \cite{Barrientos:2018cnx}. We first insert the definition of $T_{\mu\nu}$ and $\Theta_{\mu\nu}$ within Eq. \eqref{eq:tau}. After contracting the result with the metric tensor $g^{\mu\nu}$and some simplification, we have 
\begin{equation}
\begin{split}
 &{\tau^\mu}_{\nu}={T^\mu}_{\nu}-\frac{f_{Q}}{\kappa} \bigg[ 2T^{\mu\alpha} T_{\nu\alpha}-2 {\mathcal{L}^2_m} \delta^{\mu}_{\nu}\\
 &~~~~~~+2 (T+2\mathcal{L}_m) g^{\mu\lambda} \frac{\partial \mathcal{L}_m}{\partial g^{\lambda\nu}}-4 T^{\alpha\beta} g^{\mu\lambda} \frac{\partial^2 \mathcal{L}_m}{g^{\lambda\nu} g^{\alpha\beta}}\bigg].
\end{split}
\end{equation}
Then by utilizing the above relation, one can rewrite the conservation equation \eqref{eq:contau} as follows:
\begin{equation} \label{eq:divergencia}
\begin{split}
&\nabla_{\mu}^{(g)}{T^\mu}_{\nu}=-\frac{f_{Q}}{2\kappa}\partial_\nu Q+\nabla_{\mu}^{(g)} \bigg[ \frac{f_{Q}}{\kappa} \Big(2 T^{\mu\alpha} T_{\nu\alpha}-2 \mathcal{L}_m^2 \delta^{\mu}_{\nu}\\
&+4 \mathcal{L}_m g^{\mu\lambda} \frac{\partial \mathcal{L}_m}{\partial g^{\lambda\nu}}+2 T g^{\mu\lambda} \frac{\partial \mathcal{L}_m}{\partial g^{\lambda\nu}}-4 T^{\alpha\beta} g^{\mu\lambda} \frac{\partial^2 \mathcal{L}_m}{\partial g^{\lambda\nu}\partial g^{\alpha\beta}}\Big)\bigg].
\end{split}
\end{equation}
In order to simplify the above relation, we should next obtain each term on the left-hand side of this equation.
By considering the fact that
\begin{equation}\label{delrho}
\delta \rho=\frac{1}{2} \rho (u_\mu u_\nu+g_{\mu\nu})\delta g^{\mu\nu},
\end{equation}
which is comprehensively derived in equation (32) of \cite{Barrientos:2018cnx}, we find the  $\frac{\partial \mathcal{L}_m}{\partial g^{\lambda\nu}}$ and $\frac{\partial^2 \mathcal{L}_m}{\partial g^{\lambda\nu}\partial g^{\alpha\beta}}$ terms as  
\begin{equation}\label{eq1}
g^{\mu\lambda}\frac{\partial \mathcal{L}_m}{\partial g^{\lambda\nu}}=\frac{1}{2}\rho \frac{d{\mathcal{L}_m}}{d\rho}(u^\mu u_\nu +\delta^{\mu}_{\nu}),
\end{equation} 
and
\begin{equation}
\begin{split}\label{eq2}
&g^{\mu\lambda}\frac{\partial^2 \mathcal{L}_m}{\partial g^{\lambda\nu}\partial g^{\alpha\beta}}=\\
& \frac{1}{4} \rho\Big(g_{\alpha\beta}+u_\alpha u_\beta\Big)\Big(u^\mu u_\nu +\delta^{\mu}_{\nu}\Big)\bigg[ \frac{d{\mathcal{L}_m}}{d\rho}+\rho\frac{d^2 \mathcal{L}_m}{d \rho^2}\bigg],
\end{split}
\end{equation}
respectively. On the other hand, $T_{\mu\nu}$ can be written as \cite{roshan2016energy}
\begin{equation}\label{Trho}
T_{\mu\nu}= -\rho u_\mu u_\nu\frac{d \mathcal{L}_m}{d\rho}+g_{\mu\nu}\left(\mathcal{L}_m-\rho\frac{d \mathcal{L}_m}{d\rho}\right),
\end{equation}
and then $Q$ as
\begin{equation}
Q=-6\rho \mathcal{L}_m \frac{d \mathcal{L}_m}{d\rho}+3\rho^2 \left( \frac{d \mathcal{L}_m}{d\rho}\right)^2 +4 {\mathcal{L}^2_m}.
\end{equation}
Using these relations, we simplify the other terms in Eq. \eqref{eq:divergencia}, i.e., $T^{\alpha\mu} T_{\nu\alpha}$, $\partial_\nu Q$\footnote{Note, for deriving the  $\partial_\nu Q$ term, we use $\partial_\nu Q= \frac{\partial Q}{\partial \rho}\partial_\nu \rho$.}, and $T^{\alpha\beta} (u_\alpha u_\beta+g_{\alpha\beta})$, as follows:
\begin{equation}
\begin{split}
T^{\alpha\mu} T_{\nu\alpha}=&\delta^{\mu}_{\nu} {\mathcal{L}^2_m} \bigg[ \rho^2 \Big(\frac{d \mathcal{L}_m}{d\rho}\Big)^2 \\&-2\rho \mathcal{L}_m \frac{d \mathcal{L}_m}{d\rho}\bigg] \big(u^\mu u_\nu +\delta^{\mu}_{\nu}\big),
\end{split}
\end{equation}
\begin{equation}\label{parq}
\begin{split}
\partial_\nu Q =& \bigg(2\mathcal{L}_m \frac{d \mathcal{L}_m}{d\rho} +6\rho^2 \frac{d \mathcal{L}_m}{d\rho} \frac{d^2 \mathcal{L}_m }{d\rho^2}\\&-6\rho\frac{d^2 \mathcal{L}_m }{d\rho^2}\mathcal{L}_m \bigg)\partial_\nu \rho,
\end{split}
\end{equation}
\begin{equation}\label{tug}
T^{\alpha\beta} (u_\alpha u_\beta+g_{\alpha\beta})= 3\Big(\mathcal{L}_m-\rho \frac{d \mathcal{L}_m}{d\rho}\Big).
\end{equation}
Therefore, considering the mathematical definition of $u^\nu\nabla_{\nu}^{(g)}u^\mu$, i.e., 
\begin{equation}
u^\nu\nabla_{\nu}^{(g)}u^\mu=\frac{d^2x^\mu}{ds^2}+\hat{\Gamma}^\mu\,_{\lambda\nu}\frac{dx^\lambda}{ds}\frac{dx^\nu}{ds},
\end{equation}
one can rewrite Eq.\eqref{eq:divergencia} as 
\begin{equation}\label{eq:div1}
\begin{split}
\frac{d^2x^\mu}{ds^2}+\hat{\Gamma}^\mu\,_{\lambda\nu}\frac{dx^\lambda}{ds}\frac{dx^\nu}{ds}= F^{\mu},
\end{split}
\end{equation}
after some manipulation and simplification. Here, the fifth force $F^{\mu}$ is of the form
\begin{equation}\label{force}
\begin{split}
F^{\mu}=\frac{f_{Q}}{2 \kappa} \frac{1}{\mathcal{X}} &\bigg[4 \frac{d\mathcal{L}_m}{d\rho}\Big(\mathcal{L}_m-\rho \frac{d\mathcal{L}_m}{d\rho}\Big)\bigg] g^{\mu\nu} \nabla_{\nu}^{(g)}\ln \rho\\
&-\big(g^{\mu\nu} +u^\mu u^\nu\big)\nabla_{\nu}^{(g)} \left(\ln \mathcal{X} \right),
\end{split}
\end{equation}
in which
 \begin{equation}
 \begin{split}
\mathcal{X}=\frac{f_{Q}}{2 \kappa}&\bigg[-2 \mathcal{L}_m \frac{d \mathcal{L}_m}{d\rho}+4 \rho \Big(\frac{d \mathcal{L}_m}{d\rho}\Big)^2 \\
&-6 \rho \frac{d^2 \mathcal{L}_m}{d\rho^2}\Big(\mathcal{L}_m-\rho \frac{d \mathcal{L}_m}{d\rho}\Big)\bigg]+\frac{d\mathcal{L}_m}{d\rho}.
\end{split}
\end{equation}
Eq. \eqref{eq:div1} represents the trajectory of a test particle in this  theory. In GR, for a freely falling particle, the right-hand side is zero, i.e., $F^{\mu}=0$. However, in Palatini EMSG, we see that $F^{\mu}$ is nonzero and appear as a force, namely the fifth force. The particle equation of motion Eq. \eqref{eq:div1} is written in terms of connections of  the metric $g_{\mu\nu}$ and not the ``independent" connections $\Gamma^{\alpha}_{\mu\nu}$. It is necessary to recall that we assumed that the matter Lagrangian does not depend on $\Gamma^{\alpha}_{\mu\nu}$. This directly means that the Levi-Civita connection of the metric is used to define the parallel transport, and accordingly it should also appear in the geodesic equation.

In order to compare our result with the corresponding one in the metric-affine $f(R, T)$ theories, see equation (38) in \cite{Barrientos:2018cnx}, let us study the simplest case, i.e., the dust fluid, in which $p=0$ and $\mathcal{L}_m=-\rho$. For this case, Eq. \eqref{force} reduces to 
\begin{equation} \label{eq:forcep1}
F^\mu_{\text{dust}}=-(g^{\mu\nu} + u^\mu u^\nu)\partial_{\nu}\ln \left(\frac{f_{Q}}{\kappa}\rho-1\right).
\end{equation}

It is seen that the extra force depends on $f_{Q}$, while the corresponding force in the $f(R, T)$ theories is a function of $f_T$. Therefore, it seems that any kind of dependency of function $f$ on the energy-momentum content of the fluid (at least the linear and squared ones examined in \cite{Barrientos:2018cnx} and the current work, respectively) can induce an extra force. It is interesting to mention that the fifth force \eqref{eq:forcep1} in perpendicular to $u_{\mu}$, namely $F_{\mu}u^{\mu}=0$. This means that the fifth force does not have any component tangent to the four velocity vector. Of course, the spacial part of the force $\mathbf{F}$, in principle, has component tangent to the spacial velocity $\mathbf{v}$.

As expected, when $f_Q=0$, i.e., 
when $f$ is only a function of the Ricci scalar, our results reduce to those of the $f(R)$ gravitational models in the Palatini formalism.
It is worth mentioning that by adopting the standard Lagrangian density $\mathcal{L}_m=p$, one can easily show that this fifth force is zero for the dust fluid. This difference between the fifth forces for the dust fluid in two cases $\mathcal{L}_m=p$ and $\mathcal{L}_m=-\rho$ reveals that these two Lagrangian densities, like the similar case in the $f(R)$ gravity, are not equivalent, for a more detailed discussion on this issue see \cite{Faraoni:2009rk} and references therein.

It is worth mentioning that the existence of the fifth force can be considered as a direct consequence of the energy-momentum non-conservation. As we already mentioned, this causes consequences that might rule out the theory. We see that the fifth force $F^{\mu}$ depends on the internal properties of the moving object. This directly violates Einstein's equivalence principle (EEP). On the other hand, EEP is well constrained in the solar system local tests \cite{Will:2005va}. Therefore, the correction terms in EMSG should be small enough to pass the tests. We recall that the main motivation behind EMSG is to resolve the Big Bang singularity \cite{roshan2016energy}. Of course, there is no constraint or observational evidence on EEP at the early universe where EMSG's effects are not negligible. From this perspective, EMSG would remain viable at the cosmological scale. In the next section, we explore the weak field limit of Palatini EMSG to ensure the viability of the model.

As a final remark in this subsection, let us obtain the null geodesic equation in Palatini EMSG. To do so, we can use the conservation equation of a null dust fluid \cite{Sotiriou:2008it}.
In this case, the energy-momentum tensor is given by 
\begin{equation}\label{nulldust}
 T_{\mu\nu}=\rho u_\mu u_\nu. 
\end{equation}
Regarding the null condition $u_\alpha u^\alpha=0$, it is easy to show that for the null dust fluid, $Q$ vanishes. Here, $u_\alpha$ is the four-velocity of the null fluid element.  
Therefore, for this fluid, considering Eqs. \eqref{eq1}-\eqref{eq2} and using this fact that $Q=0$, one can simplify Eq. \eqref{eq:divergencia} as
 \begin{align}\label{eq3}
\nabla_{\mu}^{(g)}{T^\mu}_{\nu}=0.
\end{align}
This relation obviously illustrates that for the massless dust fluid, $T_{\mu\nu}$ is conserved in Palatini EMSG. Now, by utilizing Eq. \eqref{nulldust}, we can obtain the left-hand side of the above relation. So, after some manipulation, we arrive at
\begin{equation}
u^\mu \nabla_\mu^{(g)} u_\nu=-\big(\nabla_\mu^{(g)} u^\mu+\frac{1}{\rho} u^\mu \partial_\mu \rho\big) u_\nu,
\end{equation}
as the definition of the null geodesic curve. So, as expected from Eq. \eqref{eq3}, unlike the previous case, the null geodesic equation does not change in Palatini EMSG.

Before closing this section let us mention that deriving the modified version of the geodesic deviation equation and the Raychaudhuri’s equation in the context of EMSG would be interesting. These equations will provide a more complete picture on the particle motion in EMSG. For similar works in the context of other modified gravity theories we refer the reader to \cite{Shojai:2008zz}-\cite{Darabi:2014dla}. This issue is comprehensive enough to be considered as the subject for future studies on EMSG.
\section{Weak-field limit in Palatini EMSG} \label{sec:wf}

By weak field limit we mean that the characteristic velocity of the underlying system is small compared to the speed of light, and gravity is weak everywhere. In this situation one can ignore the relativistic effects. It is convenient to assume that the background metric  is described by the Minkowski spacetime. One can easily verify that the Minkowski metric is a trivial solution for EMSG's field equations in the vacuum. Now, a self-gravitating system can be considered as a perturbation to the background. So, we consider that the perturbed physical and auxiliary metrics are expressed by

\begin{equation}\label{g}
g_{\mu\nu} \approx \eta_{\mu\nu} + t_{\mu\nu},
\end{equation}
and 
\begin{equation}\label{hh}
h_{\mu\nu} \approx \eta_{\mu\nu}+\bar{t}_{\mu\nu},
\end{equation}
respectively.
Here, $t_{\mu\nu}$ and $\bar{t}_{\mu\nu}$ are the first order perturbations, i.e., $|t_{\mu\nu}|\ll 1$ and $|\bar{t}_{\mu\nu}|\ll 1$. 
By considering the standard form of the perturbed metric in the Newtonian limit as
\begin{equation}
h_{\mu\nu}=-(1+2\bar{\phi})dt^2+(1+2\bar{\psi})(dx^2+dy^2+dz^2)
\end{equation}
in which $\bar{\phi}$ and $\bar{\psi}$ are two scaler fields, 
one can conclude that ${\bar{t}}_{00}=-2\bar{\phi}$ and ${\bar{t}}_{ij}=2\bar{\psi}\delta_{ij}$. Here, $\delta_{ij}$ is the Kronecker delta.

Hereafter, we consider that every quantity $M^{(0)}$ is perturbed as $M=M^{(0)}+\delta M$ in which the ``${(0)}$" index denotes the non-perturbed background quantities and $\delta M$ is a very small quantity and represents the first order perturbation so that $|\delta M/M^{(0)}|\ll 1$. 
Note that in the perturbation theory, we have $\delta {M^\mu}_\nu\approx \eta^{\mu\alpha}\delta M_{\alpha\nu}$.
As we know, there is no mass/energy distribution in the Minkowskian background, So the perturbed material quantities reduce to $Q=\delta Q$, $T_{\mu\nu}=\delta T_{\mu\nu}$ and $\Theta_{\mu\nu}=\delta \Theta_{\mu\nu}$. For convenience, let us gather the perturbed gravitational quantities to the required order in the following 
\begin{subequations}
\begin{align}
\label{e2}
& R=R^{(0)}+\delta R+\cdots,\\
&{R^\mu}_{\nu}={{R^\mu}_{\nu}}^{(0)}+\delta {R^\mu}_{\nu}+\cdots,\\
\label{e4}
& f_R=f_{R}^{(0)}+f_{RR}^{(0)} \delta R+f_{RQ}^{(0)} \delta Q+\cdots ,\\
\label{e5}
& f_Q=f_{Q}^{(0)}+f_{QQ}^{(0)} \delta Q+f_{RQ}^{(0)} \delta R+\cdots ,\\
\label{e6}
& f=f^{(0)}+f_R^{(0)} \delta R+f_Q^{(0)} \delta Q+\cdots.
\end{align}
\end{subequations}
It should be recalled that $R^{(0)}$ and ${{{R}^\mu}_{\nu}}^{(0)}$ are zero in the background.
 
As a first step toward obtaining the perturbed field equations, we derive $\delta R$ by perturbing the trace equation \eqref{eq:metric} as
\begin{equation}\label{293}
\delta R=-\frac{\kappa \delta T+f_Q^{(0)} \big( 2\delta Q-\delta \Theta\big)}{f_R^{(0)}},
\end{equation}
after using Eqs. \eqref{e2}-\eqref{e6}.
We now substitute the perturbed quantities $Q$, $T_{\mu\nu}$, $\Theta_{\mu\nu}$, and \eqref{e2}-\eqref{e6} into the time-time component of Ricci tensor \eqref{eq:Riccieqs}. After utilizing Eq. \eqref{293} and keeping the first-order terms, we arrive at
\begin{equation} \label{eq:R00}
\delta R^{0}_{0}=\frac{\kappa}{f_{R}^{(0)2}} \bigg[ \delta T^{0}_{0}-\frac{1}{2} \delta T+\frac{f_{Q}^{(0)}}{2 \kappa}\Big(\delta \Theta-2\delta \Theta^{0}_{0}-\delta Q\Big)\bigg].
\end{equation}
We will see that for our purpose, computing this component is sufficient and we do not need to obtain the rest components of $\delta R^{\mu}_{\nu}$.

For finding the right-hand side of the above relation, let us briefly introduce the perturbed matter quantities. Taking the energy-momentum tensor of the perfect fluid \eqref{eq:pftumunu}, one can show that
\begin{equation}\label{pfQ}
\delta T^{0}_{0}=-p~~,~~\delta T=-\rho~~,~~\delta Q=\delta \Theta=\delta \Theta^{0}_{0}=\rho^2, 
\end{equation}
Therefore, by using these quantities, we can rewrite the right-hand side of Eq. \eqref{eq:R00} in terms of pressure and energy density. 
It is worth mentioning that in the weak-field regime, one can neglect the contribution of pressure compared with the energy density. In fact, in this limit, we have $ |T_{jk}|\ll| T_{0j}|\ll| T_{00}|$.

Here, in order to construct the left-hand side of Eq. \eqref{eq:R00} from the auxiliary perturbation metric $\bar{t}_{\mu\nu}$, we first perturb Eq. \eqref{gamma} by using $h_{\mu\nu} \approx \eta_{\mu\nu}+\bar{t}_{\mu\nu}$, and find the first-order Christoffel symbols. Then by inserting $\delta \Gamma^{\lambda}_{\mu\nu}$ within the definition of the Ricci tensor, after some simplification, we find the following first order equation
\begin{equation}\label{deltaR}
\begin{split}
\delta R_{\mu\nu}=-\frac{1}{2} \Big(\Box\bar{t}_{\mu\nu}-\partial_\lambda \partial_\mu \bar{t}^{\lambda}_{\nu}-\partial_\lambda \partial_\nu \bar{t}^{\lambda}_\mu+\partial_\mu \partial_\nu \bar{t}\Big),
\end{split}
\end{equation}
where $\Box=\partial^\alpha\partial_\alpha$ is the wave operator in the Minkowskian spacetime and $\bar{t}$ is the determinant of the auxiliary perturbation metric $\bar{t}_{\mu\nu}$.
By choosing the standard gauge $\partial_\lambda (\bar{t}^{\lambda}_{\mu} -\frac{\bar{t}}{2} \delta_{\mu}^{\lambda})=0$, one can show that the sum of the last three terms in Eq. \eqref{deltaR} becomes zero and this relation reduces to the following form
\begin{equation} \label{star}
\delta R^{\mu}_\nu (h)=-\frac{1}{2} \Box\bar{t}^{\mu}_{\nu}\simeq-\frac{1}{2} \nabla^2 \bar{t}^{\mu}_{\nu}.
\end{equation}
Notice that since the time derivative is smaller than the spatial derivative in the weak-field regime \cite{poisson2014gravity}, we can replace the d'Alembertian operator $\Box$ with the Laplacian operator $\nabla^2$.
Now, by substituting the time-time component of Eq. \eqref{star} into Eq. \eqref{eq:R00} and also applying Eq. \eqref{pfQ}, we finally obtain 
\begin{equation} \label{eq:poisson}
\nabla^2 \bar{\phi}=\frac{\kappa}{f_{R}^{(0) 2}}\bigg[\frac{1}{2} \rho+\frac{f_{Q}^{(0)}}{\kappa} \rho^2\bigg].
\end{equation}

This equation cannot describe the nature of the modified $f(R, Q)$ gravity in the weak-field limit. In fact, the scaler field $\bar{\phi}$ is the auxiliary Newtonian potential based on $h_{\mu\nu}$ which is not the actual spacetime. Therefore, to find the true description of the gravity in Palatini EMSG, we should find the relationship between $\bar{\phi}$ and the physical Newtonian potential $\phi$ related to $g_{\mu\nu}$. To do so, let us return to the conformal relation between $h_{\mu\nu}$ and $g_{\mu\nu}$ once again. By perturbing this relation and also using Eqs. \eqref{g} and \eqref{hh}, we find  
\begin{equation}\label{gdelta}
 t_{\mu\nu}=\frac{1}{f_{R}^{(0)}}\bigg(\bar{t}_{\mu\nu}-\eta_{\mu\nu} \delta f_{R}\bigg),
\end{equation} 
for the first-order equation. Furthermore, from the Newtonian order of the conformal relation, one can easily conclude that $f_{R}^{(0)}=1$. Although, here, we assume that $\delta f_{R}$ does not vanish and it is given by $\delta f_{R}=f_{RR}^{(0)}\delta R$ \cite{Barrientos:2018cnx}. 
Using this fact that $ t^{0}_{0}=2\phi$ and inserting Eq. \eqref{293} within Eq. \eqref{gdelta} led to the following relation
\begin{equation}
 \phi=\bar{\phi}-\frac{\kappa f_{RR}^{(0)}}{2}\bigg(\rho-\frac{f_Q^{(0)}}{\kappa}\rho^2\bigg).
\end{equation}
Finally, by regarding the above equation, we show that Eq. \eqref{eq:poisson} reduces to 
\begin{equation}
\nabla^2 \phi=\kappa\bigg(\frac{\rho}{2}+\frac{f_Q^{(0)}}{\kappa}\rho^2\bigg)+\frac{\kappa f_{RR}^{(0)}}{2}\bigg(\frac{f_Q^{(0)}}{\kappa}\rho-1\bigg)\nabla^2 \rho.
\label{pof}
\end{equation}
In fact, this equation represents the modified version of the Poisson equation in the weak-field limit of the Palatini EMSG.
Note that in this modified Poisson equation, the first term in the right-hand side, i.e., $\kappa\,\rho/2$ denotes the standard Poisson equation in the Newtonian gravity and the rest terms are new correction terms induced by Palatini EMSG.
It is seen that these correction terms containing the derivations of $f$ with respect to $R$ and $Q$, may cause significant deviation from the Newtonian gravity in the environments where those are not negligible. It is constructive to mention when $f_Q^{(0)}\rightarrow 0$, the above relation reduces to the corresponding Poisson equation in the Palatini $f(R)$ gravity, e.g., see  \cite{Dominguez:2004ds} and references therein.

Eq. \eqref{pof} can  be compared with equation (32) in \cite{kazemi2020jeans}. In this paper the weak field limit of EMSG in the metric formulation has been investigated. It is interesting that metric and Palatini versions lead to different weak field limits. In the Palatini version we have second derivatives of the matter density in the right-hand side of the modified Poisson equation. While, in the metric formulation, there is a nonlocal integral term over the whole space. It is also interesting to mention that the existence of the Laplacian of the matter density in the Poisson equation is reminiscent of the weak field limit of EiBI gravity \cite{Banados:2010ix}. This correction term may cause drastic consequences in the properties of compact objects and the gravitational collapse. In the case of Palatini $f(R)$ gravity, i.e., $f_Q^{(0)}\rightarrow 0$, it has been recently argued in \cite{Olmo:2020fri} that proper junction conditions formalism is needed to be adapted in order to have well defined compact objects. We refer the reader to \cite{Pani:2012qb} where several aspects of the existence of $\nabla^2 \rho$ in the Poisson equation have been investigated.

For a given matter distribution $\rho(\mathbf{x})$, one may solve equation \eqref{pof} to find the corresponding potential. Then the trajectory of the particles can be found using the geodesic equation written in the weak field limit, namely the modified version of the Newtonian equation of motion for a test particle in the gravitational field.

From this point of view, it would be desirable to find the weak field limit of the geodesic equation \eqref{eq:div1}. As mentioned before, in the weak field limit, we assume that the particle moves slowly. This is equivalent to considering that $|\frac{d x^i}{d s}|\ll |\frac{dx^0}{ds}|$ in this case. By utilizing this fact, Eq. \eqref{eq:div1} reduces to 
\begin{equation}
\frac{d^2x^\mu}{ds^2}+\hat{\Gamma}^{\mu}_{00}\Big(\frac{dx^0}{ds}\Big)^2= F^{\mu}.
\end{equation}
We recall that $s$ is an arbitrary affine parameter and $x^\mu(s)$ represents the particle worldline in the curved spacetime.
In the first order perturbation, one can find that $\hat{\Gamma}^{\mu (0)}_{00}=0$ and $\hat{\Gamma}^{\mu}_{00}\simeq\delta \hat{\Gamma}^{\mu}_{00}=-\frac{1}{2}\eta^{\mu\nu}\partial_\nu t_{00}$. Here, we assume that the metric is stationary. We should also mention that in the Minkowskian background, $F^{\mu (0)}=0$ and consequently, $F^{\mu}\simeq\delta F^{\mu}$. Therefore, by considering these assumptions and $t_{00}=-2\phi$, we have 
\begin{equation}
\label{geoweak}
\frac{d^2x^\mu}{ds^2}+\eta^{\mu\nu}\partial_\nu \phi \Big(\frac{dx^0}{ds}\Big)^2= \delta F^{\mu}.
\end{equation}

We now turn to obtain $\delta F^{\mu}$. To do so, let us first apply the standard Lagrangian density  $\mathcal{L}_m=p$. Here, we also use the barotropic equation of state for the matter, i.e., $p=w \rho $. For this case, we arrive at
\begin{align}
\label{F^mu}
F^{\mu}=-\big(g^{\mu\nu}+\frac{dx^{\mu}}{ds}\frac{dx^{\nu}}{ds}\big)\partial_{\nu}\ln \mathcal{X},
\end{align}
as the fifth force in which
\begin{align}
\label{X}
\mathcal{X}=\frac{w^2}{\kappa}f_{Q} \rho+w.
\end{align}
Since there is no mass/energy distribution in the Minkowskian background, the zero and first orders of Eq. \eqref{X} are given by $\mathcal{X}^{(0)}=0$ and $ \delta\mathcal{X}=w^2/\kappa f_{Q}^{(0)}\rho+w$, respectively. Therefore, one can obtain that 
\begin{align}
\delta F^{\mu}=-\big(\eta^{\mu\nu}+\frac{dx^{\mu}}{ds}\frac{dx^{\nu}}{ds}\big)\partial_{\nu}\ln \big(\frac{w^2}{\kappa}f_{Q}^{(0)}\rho+w\big),
\end{align}
We notice again that in the weak-field regime, one can neglect the contribution of pressure compared with the energy density. In fact, in this limit, we consider that $w=\textit{cte}$ which is very small, i.e., $w\ll 1$. By keeping this fact in mind, we find that $\delta F^{\mu}=0$. Therefore, for this Lagrangian density, the time and space components of Eq. \eqref{geoweak} reduce to 
\begin{align}
\frac{d^2t}{ds^2}=0,~~~~~~~~~~~~~~\frac{d^2x^i}{ds^2}=-\partial_i \phi \big(\frac{dt}{d s}\big)^2,
\end{align}
respectively. Here, $t$ is defined by $x^0\equiv t$. The first relation in the above equations revels that $dt/ds=\textit{cte}$. So, the second relation can be written in the following vectorial form
\begin{align}\label{d2x}
\frac{d^2\vec{x}}{dt^2}=-\nabla \phi.
\end{align}

As can be seen, although, the mathematical form of the above relation is quite similar to the Newtonian equation of motion for a particle in the weak field limit of GR, regarding Eq. \eqref{pof}, a particle can move in a completely different path in the weak-field regime of the Palatini EMSG. We mention that in this case, the fifth force has no role in the worldline of the particle. However, EMSG's corrections appear in the gravitational sector.

For the sake of completeness, it is also instructive to investigate the weak field limit of the geodesic equation \eqref{geoweak}  in the case of $\mathcal{L}=-\rho$. By considering Eq. \eqref{eq:forcep1}, one can find that
\begin{align}
\delta F^{\mu}_{\text{dust}}=-\big(\eta^{\mu\nu}+\frac{dx^{\mu}}{ds}\frac{dx^{\nu}}{ds}\big) \partial_{\nu}\ln \Big(\frac{f_{Q}^{(0)}}{\kappa}\rho\Big).
\end{align}
By inserting the above relation within Eq. \eqref{geoweak} and using $\partial_{\mu} f_{Q}^{(0)}=f_{QQ}^{(0)}\partial_{\mu} \delta Q$, after some manipulations, we arrive at
\begin{align}
\label{d2t}
\frac{d^2t}{ds^2}=0,
\end{align}
and
\begin{align}
\label{d2xs}
\frac{d^2\vec{x}}{ds^2}=-\nabla\phi\Big(\frac{dt}{ds}\Big)^2-\frac{\Big( f_{Q}^{(0)}+2\rho^2 f_{QQ}^{(0)}\Big)}{ f_{Q}^{(0)}}\frac{\nabla\rho}{\rho},
\end{align}
for the time and space components of Eq. \eqref{geoweak}, respectively. Then by using Eq. \eqref{d2t} and keeping in mind that $ds^2\simeq g_{00}dt^2$, we arrive at 
\begin{align}
\label{d2xt}
\frac{d^2\vec{x}}{dt^2}=-\nabla\phi+\frac{\Big( f_{Q}^{(0)}+2\rho^2 f_{QQ}^{(0)}\Big)}{ f_{Q}^{(0)}}\frac{\nabla\rho}{\rho},
\end{align}
which is the modified version of the equation of motion describing the path of a particle. In this case, in contrast to the previous one, i.e., when $\mathcal{L}_m=p$, the fifth force plays a significant role in determining the worldline of the particle. It is obvious if either $f_{Q}^{(0)}+2\rho^2 f_{QQ}^{(0)}=0$ or $\nabla\rho=0$ is established, Eq. \eqref{d2xt} reduces to Eq. \eqref{d2x}. Therefore, only under these specific conditions, a particle can fall freely on the same path in both Lagrangian densities $\mathcal{L}_m=p$ and $\mathcal{L}_m=-\rho$.

It is also important to mention that outside the matter distribution where $\rho=0$, the geodesic equation coincides with the standard case. In other words, there is no fifth force experiencing by the test particle moving within the vacuum. More interestingly, this theory does not induce any modification to vacuum metric solutions in Palatini $f(R)$ gravity. One should note that vacuum solutions in the Palatini and metric formulations of $f(R)$ gravity are effectively equivalent to ``vacuum" solutions in GR+$\Lambda$. Therefore, one may conclude that vacuum solutions in Palatini/metric EMSG are equivalent to those of in GR+$\Lambda$, like de Sitter/anti-de Sitter Schwaraschid and Kerr-de Sitter metrics. For example, the linear model investigated in \cite{roshan2016energy} does not change GR's vacuum solutions like Schwarzschild and Kerr metrics.

As the last point in this section, it should be noted that we expect the corrections in the modified Poisson equation of EMSG are negligible even in the galactic scales. In other words, there are modified gravity theories that are identical to GR at solar system experiments while showing significant deviations from GR at galactic scales. The deviations are helpful in the sense that they can alleviate the missing mass problem in the galaxies, for example see \cite{Hehl:2009es} and \cite{Moffat:2005si}. However, EMSG is by no means relevant to the dark matter problem. This theory's effects appear only at the strong field limit. In the next section, we explore the cosmological behavior of the theory.

\section{Cosmic bounce in Palatini EMSG}\label{Cosmicbounce}

As we mentioned before, EMSG deviates from GR at high curvature. Therefore, we expect different behavior for this theory in the very early universe. In the case of the metric EMSG, it has been shown that the universe does not necessarily start from a singularity and it can instead cross a bounce at which the maximum energy density of the cosmic fluid has a finite value \cite{roshan2016energy}. 

We know that the cosmic bounce can exist in the Palatini formalism of modified theories of gravity, e.g., for $f(R)$ theories see \cite{Barragan:2009sq} and references therein. Therefore, as the first cosmological consequence of Palatini EMSG, we are interested to identify the characterization of a bounce in the early time cosmology.

For our purposes, it is convenient to rewrite the Einstein tensor in terms of the physical metric $g_{\mu\nu}$. To do so, as mentioned before, we utilize the conformal transformation $h_{\mu\nu}=f_{R}g_{\mu\nu}$. After some manipulation and simplification, we express Eq. \eqref{eq:metric} in the form
\begin{equation}\label{G(g)}
\begin{split}
G_{\mu\nu}(g)=& G_{\mu\nu}(h)+\frac{1}{f_{R}}\Big(\nabla_{\mu} \nabla_{\nu} f_{R} - g_{\mu\nu}\Box f_{R}\Big)\\
&-\frac{3}{2 f_{R}^2}\Big(\partial_\mu f_{R} \partial_\nu f_{R}-\frac{1}{2}g_{\mu\nu}\partial _{\lambda}f_R\partial^{\lambda}f_R\Big),
\end{split}
\end{equation}
in which $G_{\mu\nu}(g)$ is the Einstein tensor obtained from the physical metric. It should be mentioned that, here, the covariant derivatives are written in terms of $g_{\mu\nu}$. We assume that the distribution of matter and energy in the universe can be described with the perfect fluid and consequently, the metric of the space-time is given by the Friedmann-Robertson-Walker (FRW) metric 
\begin{eqnarray}
ds^2=-dt^2+ a(t)^2\Big(&\frac{dr^2}{1-k r^2}+r^2d\Omega^2\Big),
\end{eqnarray}
where $a(t)$ is the cosmological scale factor and $k=0,\pm 1$ represents the spatial curvature. We assume that the equation of state of the cosmic fluid is $p=w\rho$. Then one can readily verify that $Q=\left(3w^2+1\right)\rho^2$. By considering the above assumptions, and inserting Eq. \eqref{G(g)} into Eq. \eqref{eq:Einstein}, we finally derive time-time and space-space components of Eq. \eqref{eq:Einstein} as 

\begin{widetext}
\begin{equation}\label{H2}
H^2= 2\bigg(\frac{2 \rho  \Big(\alpha  \rho  f_Q +\kappa\Big)+\Pi_1}{3 f_R}-\frac{2 k}{a^2}\bigg)\Bigg(\frac{\Gamma_1 \Pi_2}{2 \Gamma_1 \Pi_2-3  (w+1)\rho \Gamma_2  f_{RR}\Big(   (3 w+1) \rho f_Q+\kappa \Big)}\Bigg)^2,
\end{equation}
\begin{equation}\label{Hprim}
\begin{split}
& 4 \dot{H} \bigg(2-\frac{3}{\Gamma_1 \Pi_2} (w+1)  \rho \Gamma_2 \Gamma_3 f_{RR}\bigg)+\frac{4 k}{a^2}-\frac{2}{f_R}\Big(\Pi_1-2 \kappa w \rho \Big)\\&+3H^2\Bigg(4+\frac{1}{\Gamma_1^3\Pi_2^3}\bigg[ 3 (w+1)^2\rho ^2\Gamma_1 \Gamma_2^2 \Gamma_3^2 f_R \Big(4 f_R^2 f_{RRR}-3 f_R f_{RR}^2+3 R f_{RR}^3\Big)+4 (w+1) \rho \Phi \Gamma_3  \Pi_2^2  f_{RR}\bigg]\Bigg)=0, 
\end{split}
\end{equation}
\end{widetext}
receptively.  Here, $\alpha=(w+1)(3w+1)$ and $H=\frac{\dot{a}}{a}$ where the dot sign stands for a derivative with respect to time. For simplifying these relations, we also introduce the following functions
\begin{equation}
\Gamma_1=\kappa+(\alpha +4 w) \rho f_Q +2 \alpha \left(3
   w^2+1\right) \rho ^3 f_{QQ},
\end{equation}
\begin{equation}
\Gamma_3=\kappa+\left(3w+1\right)\rho f_Q,
\end{equation}
\begin{widetext}
\begin{equation}
\Gamma_2=\kappa\left(1-3w\right)+2 \left(\alpha -6 w^2-2\right) \rho f_{Q}+2\alpha \left(3w^2+1\right) \rho ^3 f_{QQ},
\end{equation}
\begin{equation}
\Gamma_4=3 \left(\alpha -6 w^2-2\right) f_Q^2 +\left(3
   w^2+1\right) \left(9 \alpha-24 w^2-8\right)\rho ^2 f_{Q} f_{QQ}+2 \alpha \left(3 w^2+1\right)^2 \rho ^4 \left(f_{QQ}^2+f_Q f_{QQQ}\right),
\end{equation}
\begin{equation}
\Gamma_5= \left(2 \alpha -21 w^2-3\right)f_{Q}+\left(3 w^2+1\right)\rho ^2\bigg(3 \left(2 \alpha -7 w^2-1\right)f_{QQ}+2 \alpha \left(3 w^2+1\right)
    \rho ^2 f_{QQQ}\bigg),
\end{equation}
\begin{equation}
\Gamma_6= (\alpha +4 w)f_{Q}+4\left(3 w^2+1\right) \rho ^2  \bigg(2(\alpha +w)f_{QQ}+\alpha \left(3 w^2+1\right)\rho ^2 f_{QQQ} \bigg),
\end{equation}
\begin{equation}
\Phi=-2 \Gamma_1^2\Gamma_2-3(w+1)\rho\Gamma_2\Gamma_3\Gamma_6 +3(w+1) \Gamma_1 \left(\kappa ^2 (1-3 w)+2 \kappa  \rho \Gamma_5 +2 (3 w+1)\rho ^2 \Gamma_4\right),
\end{equation}
\end{widetext}
\begin{equation}
\Pi_1=R f_R-f,
\end{equation}
\begin{equation}
\label{Pi_2}
\Pi_2=f_R\left(f_R-R f_{RR}\right),
\end{equation}

So, Eqs. \eqref{H2} and \eqref{Hprim} are enough to find the unknown functions, i.e. $a(t)$ and $\rho(t)$. It should be mentioned that in order to simplify these equations, we applied Eq. \eqref{eq:connection} and the conservation relation
\begin{equation}\label{pcontau}
\dot{\rho}=-\frac{3 H(w+1)\Big(\kappa\rho+(1+3w)\rho^2 f_{Q}\Big)}{\kappa+(\alpha+4w)\rho f_{Q}+2 \alpha(3w^2+1)\rho^3f_{QQ}}.
\end{equation}
It is instructive to note that by setting $f_Q$, $f_{QQ}$, and $f_{QQQ}$ to zero, these relations reduce to corresponding equations in Palatini $f(R)$ theories introduced in \cite{Barragan:2009sq}. As expected, the FRW equations in GR are also reproduced in the linear limit $f \rightarrow R$.

Now, we are in a position to study in what conditions a regular bounce can happen in Palatini EMSG. In the following subsections, we first introduce the general bouncing conditions. Then, by choosing a toy model for $f(R,Q)$, we examine these conditions one by one and investigate the existence of a cosmic bounce in the very early universe.

\subsection{General bouncing conditions}\label{General bouncing conditions}
In this subsection, we introduce the bouncing conditions and characterize possible cosmic bounces. Henceforth, for the sake of simplicity, we consider a flat geometry and set $k=0$ in Eqs. \eqref{H2} and \eqref{Hprim}. In the below, for convenience, we categorize these conditions:

\textit{i)}  $\dot{a}_{\text{B}}=0$
is known as the first necessary condition for the existence of a bounce. i.e., the scale factor can reach an extremum where a cosmic bounce may exist. Here, the index ``B" represents the bounce. Under this condition, we have $H_{\text{B}}=0$. As seen from the first Friedmann equation \eqref{H2}, there are three cases for which $H$ can be zero at $\rho=\rho_{\text{B}}$, i.e., when $\Delta_{\text{B}}=0$, $\Gamma_{1\text{B}}=0$, or $\Pi_{2\text{B}}=0$. Here, $\Delta$ is defined as $\Delta=2 \rho\big(\alpha  \rho  f_Q +\kappa\big)+\Pi_1$. We will separately examine each case in the following.

\textit{ii)} The second condition for having a bounce is that this extremum should be a minimum. Therefore, we put $\ddot{a}_{\text{B}}>0$. As $H=0$ at the bounce, this condition reduces to $\dot{H}_{\text{B}}>0$. 
Using Eq. \eqref{Hprim}, one can obtain 
\begin{equation}\label{cond-2}
\dot{H}_{\text{B}}=\Big(\frac{\left(\Pi_1-2 \kappa w \rho\right)\Gamma_1 \Pi_2}{f_R\big(4\Gamma_1 \Pi_2-6(w+1)\rho \Gamma_2 \Gamma_3 f_{RR}\big)}\Big)_B>0,
\end{equation}
as the second bouncing condition.

\textit{iii)} In order to have a maximum energy density at the bounce point, the next condition is that $\ddot{\rho}_{\text{B}}<0$.
Since $\dot{H}$ should be positive at this point, as expressed in the previous part, the third condition can be written as 
\begin{equation}\label{cond-3}
\Big(\frac{(w+1)\Big(\kappa\rho+(1+3w)\rho^2 f_{Q}\Big)}{\kappa+(\alpha+4w)\rho f_{Q}+2 \alpha(3w^2+1)\rho^ 3f_{QQ}}\Big)_{\text{B}}>0,
\end{equation}
after applying Eq. \eqref{pcontau}. This is true provided that the denominator does not vanish at the bounce. Furthermore, to have a regular bounce and to avoid singularity in the continuity equation, the denominator should not vanish in the interval $\rho\in (0, \rho_{\text{B}})$, or its root coincides with that of the term $\Big(\kappa+(1+3w)\rho f_{Q}\Big)$ in the numerator. The singularity should be carefully checked in other governing equations. Regarding the complicated nature of the equations for a general $f(R, Q)$ function, it is difficult to find a single condition to remove singularity in all the governing equations. However, as we shall show, for our toy model this issue can be straightforwardly addressed.

\textit{iv)} As we know, $H$ is a real function, so $H^2$ must be positive. To fulfill this condition, by considering the Eq. \eqref{H2} and also ignoring $2k/a^2$, we obtain that $\Delta f_R>0$ within the interval $0<\rho<\rho_{\text{B}}$.

\textit{v)} As the last bouncing condition, it should be recalled that to avoid having a singularity at the beginning of the universe, the minimum value of the scale factor should not be zero, i.e., $a_{\text{B}}\neq 0$. In fact, it means that the Hubble parameter $H$ remains finite in the very early universe. Therefore, by considering Eq. \eqref{H2}, the fifth condition is that
\begin{equation}
\Big(2 \Gamma_1 \Pi_2-3  (w+1)\rho \Gamma_2  f_{RR}\big((3 w+1) \rho f_Q+\kappa \big)\Big)_{\text{B}}\neq 0.
\end{equation}
Moreover, in order to have a regular bounce and to avoid singularity in the Friedman equation, the roots of this relation should not be in the interval $\rho\in (0, \rho_{\text{B}})$.  

To sum up, we have defined five bouncing conditions. When all of these conditions are satisfied and also the root of Eq. \eqref{H2}, $\rho_{\text{root}}$, be a real value, $\rho_{\text{root}}$ can represent a physical cosmic bounce, i.e., $\rho_{\text{root}}=\rho_{\text{B}}$.
We next return to specify the cases under which $H$ can be zero. According to the discussion in the first part, there are three cases that satisfy this condition. In the following,  for each case, we examine the rest bouncing conditions.

\subsubsection{Case A: $\Pi_{2{\text{B}}}=0$} 
According to the definition \eqref{Pi_2}, this case means that $f_R$ or $f_{R}-R f_{RR}$ is zero at $\rho=\rho_{\text{B}}$. It is worth mentioning as the bounce happens at some high curvature in the very early universe, it is expected that $R_{\text{B}}\sim R_{\text{P}}$ where $R_{\text{P}}$ is the Planck curvature written in terms of the Planck length as $R_{\text{P}}\sim \ell_{\text{P}}^{-2}$.

For the case $f_{R}=0$  (at the bounce), by substituting $f_{R}=0$ into Eq. (\ref{eq:traceEq}), after some algebra, one can obtain that 
\begin{equation}
f_{\text{B}}=\frac{1}{2} \rho \Big(\alpha \rho f_{Q}+\kappa(1-3 w)\Big)_{\text{B}},
\end{equation}
\begin{equation}
\Delta_{\text{B}}=\frac{3}{2} \rho \Big(\alpha\rho f_{Q} +\kappa (w+1)\Big)_{\text{B}}.
\end{equation}
Now, by inserting $f_{\text{B}}$ within the second condition \eqref{cond-2}, we find that
\begin{equation}
\dot{H}_{\text{B}}=\Big(-\frac{\left(\kappa(w+1)\rho+\alpha \rho^2 f_{Q}\right)R\Gamma_{1}}{12(w+1)\rho\Gamma_{2}\Gamma_{3}}\Big)_{\text{B}}>0.
\end{equation}
Also, since the third condition is a function of $\rho$, $f_{Q}$, and $f_{QQ}$, the mathematical appearance of Eq. \eqref{cond-3} does not change in this case.  Moreover, in this case, the fourth and fifth conditions can be established. Hence, $f_R=0$ (at the bounce) can be one of the possible case for creating a cosmic bounce.

In the case $\big(f_{R}-R f_{RR}\big)_{\text{B}}=0$, both $H$ and $\dot{H}$ are equal to zero. So, the second condition is not satisfied and consequently, this case cannot create a bounce.

\subsubsection{ Case B: $\Gamma_{1{\text{B}}}=0$}

For this case, one can show that $\dot{H}_{\text{B}}$ is zero. Therefore, we conclude that this case like the latter one in the previous part, cannot induce a cosmological bounce.

\subsubsection{ Case C: $\Delta_{\text{B}}=0$}
This condition means that at the bounce we have 
\begin{equation}
f_Q=-\frac{\Pi_1+2\kappa \rho}{2\alpha\rho^2}.
\end{equation}
By inserting this relation within Eq. \eqref{eq:traceEq}, after some simplification, we obtain that
\begin{equation}
 f_{\text{B}}=\Big(\frac{1}{3}R f_R-2\kappa w \rho\Big)_{\text{B}}.
 \end{equation}
For this value, the second bouncing condition reduces to
\begin{equation}
\dot{H}_{\text{B}}=\Big(\frac{2 R\Gamma_1\Pi_2}{12\Gamma_1\Pi_2-18\left(w+1\right)\rho\Gamma_2\Gamma_3 f_{RR}}\Big)_{\text{B}}>0,
\end{equation}
and the condition $\ddot{\rho}_{\text{B}}<0$ is given by
\begin{equation}
\begin{split}
&\alpha \rho R f_{R}\bigg[(3w^2+8w+1)R f_R+3\Big((3w^2+5w)\kappa\rho\\
&-2\alpha(3w+1)(3w^2+1)\rho^4 f_{QQ}\Big)(w+1)\bigg]^{-1}_{\text{B}}>0.
\end{split}
\end{equation}

In this case, the rest of the bouncing conditions, i.e., the fourth and fifth conditions, can be satisfied. Therefore, the current case can also be considered as one of the possible situations where a non-singular solution exists.

We have so far introduced the general conditions for having a bouncing solution. In fact, these results are valid for any $f(R,Q)$ theory near the bounce. As shown, without fixing $f(R,Q)$, the above relations are quite complicated. So, in order to investigate these conditions and gain more intuition, we will choose a simple and also helpful model for the $f(R,Q)$ theory in the very early universe and obtain the cosmic bounce for each case in the next subsection.

As the last part of this subsection, let us recall the energy conditions.
For the perfect cosmic fluid with $p=w\rho$, it is shown that the general forms of the energy conditions are given by 
\begin{subequations}
\begin{align}
\label{NEC}
&\rho(1+w)\geq 0,\\
&\rho\geq 0, \text{and}\, \rho(1+w)\geq 0,\\
\label{SEC}
&\rho(1+w)\geq 0, \text{and}\, \rho(1+3w)\geq 0,
\end{align}
\end{subequations}
which are respectively known as the null energy condition (NEC), weak energy condition (WEC), and strong energy condition (SEC) \cite{Novello:2008ra}.
We know that one or more energy conditions are violated in the bouncing models, for instance see \cite{Novello:2008ra} and references therein. It should be added by regarding that WEC is a necessary condition for establishing the dominant energy condition (DEC), we do not consider DEC here.
In order to investigate these conditions in Palatini EMSG, we should obtain $\rho$  in terms of $a$ from Eqs. \eqref{H2} and \eqref{Hprim}. After inserting this relation within the above conditions and considering  $\dot{a}=0$ and $\ddot{a}>0$, we can examine these conditions in this formalism. To do so, we should first choose a toy model for $f(R, Q)$. We will postpone this issue to the next subsection and show how these conditions restrict the allowed value of $w$.

\subsection{A toy model}

We launch our calculation by choosing a toy model of $f(R, Q)$. In this model, $f$ is a linear and quadratic function of $Q$ and $R$, respectively. It takes the following form  
\begin{equation}
f(R, Q)=R+\beta R^2+\eta Q,
\end{equation}
in which $\beta$ and $\eta$ are coupling constants that their values can be determined by observational constraints. 
It is necessary to mention that the Palatini $f(R)$ model $f(R)=R+\beta R^2$ already leads to bouncing cosmology \cite{Barragan:2009sq}. Therefore it is natural to expect the existence of bounce in our model. Notice that our aim here is just to present the simplest toy model in EMSG.
It is obvious that in the solar system local tests these parameters should satisfy the constraint $|\eta|\ll|R/Q|$ and $|\beta|\ll 1/|R|$. It should be mentioned that both $\beta$ and $\eta$ can be positive or negative. The coefficient of the linear term $R$ is a unity to ensure that at low curvature regime this model recovers GR, i.e., $\lim_{R\rightarrow 0}f_R=1$.
For this model, using the trace equation \eqref{eq:traceEq} one can find that 
\begin{equation}
R=(3w-1)\Big((1-w)\eta \rho^2-\kappa\rho\Big).
\end{equation}
It should be added since our aim is to describe the very early universe, there is no need to consider the role of the cosmological constant $\Lambda$ in our toy model.
However, we will bring it back when we attempt to estimate the magnitude of the coupling constants $\eta$ and $\beta$. It is necessary to mention that in the metric formulation the existence of the cosmological constant can play an essential role for the bounce. For example, in the simple linear model investigated in \cite{roshan2016energy}, without $\Lambda$ one cannot achieve a non-singular solution. Therefore, interestingly, the cosmological constant in that model plays essential roles in both the early universe (for addressing the singularity problem) and the late time universe (for explaining the cosmic speed-up). Such behavior might happen in the Palatini version. However, we omit such a possibility and focus on bounces without $\Lambda$.

Before introducing the bouncing conditions for this model, let us rewrite Eqs. \eqref{H2}, \eqref{Hprim}, and \eqref{pcontau} in a dimensionless form. The dimensionless form of the equations is helpful for our subsequent calculations.
To achieve this goal, we introduce the dimensionless quantities as follows
\begin{equation}
\begin{split}
&\rho_{\text{d}}=\frac{\mid \eta \mid}{\kappa}\rho=\text{sign}(\eta)\frac{\eta}{\kappa}\rho,~~~~t_{\text{d}}=\frac{\kappa}{\sqrt{\mid \eta \mid}}t,\\&
\beta_{\text{s}}=\frac{\kappa^2}{\eta}\beta,~~~\rho_{\text{s}}=\frac{\rho}{\rho_{\text{B}}}=\frac{\rho_{\text{d}}}{\rho_{\text{Bd}}}, ~~~~H_{\text{d}}=\frac{\sqrt{\mid \eta \mid}}{\kappa} H.
\end{split}
\end{equation}
Using these new variables, and after some simplification, Eq. \eqref{H2} takes the following form
\begin{widetext}
\begin{equation}\label{Hd}
\begin{split}
 H^2_\text{d}=&-\frac{Y^2}{6 X^2}\rho_{\text{d}}\bigg(2(3w-1)\Big(1+(w-1)\text{sign}(\eta)\rho_{\text{d}}\Big)\text{sign}(\eta)\beta_{\text{s}}\rho_{\text{d}}-1\bigg)\times\\
&\bigg(2+\Big[\lambda+(9w^2-6w+1)\beta_{\text{s}}+\gamma(3w-1)\Big(2+(w-1)\text{sign}(\eta)\rho_{\text{d}}\Big)\text{sign}(\eta)\beta_{\text{s}}\rho_{\text{d}}\Big]\text{sign}(\eta)\rho_{\text{d}}\bigg),
\end{split}
\end{equation}
in which $\lambda=\alpha+4w$, $\gamma=(3w-1)(w-1)$ and 
\begin{equation}
X=1+\bigg[\lambda+(9w^2-1)\beta_{\text{s}}+\Big(9w+3+4(3w^2+2w+1)\text{sign}(\eta)\rho_{\text{d}}\Big)\gamma\text{sign}(\eta)\beta_{\text{s}}\rho_{\text{d}}\bigg]\text{sign}(\eta)\rho_{\text{d}},
\end{equation}
\begin{equation}
Y=1+\lambda\text{sign}(\eta)\rho_{\text{d}}.
\end{equation}
\end{widetext}
By applying the dimensionless parameters and functions, one can also obtain the dimensionless form of the second Friedmann equation \eqref{Hprim} as follows 
\begin{widetext}
\begin{equation}
\begin{split}
 8 & H'_{\text{d}}\Bigg(1-2\gamma\beta_{\text{s}}\rho_{\text{d}}^2+\frac{1}{Y^3}\Big[(9w^2-1)\text{sign}(\eta)\beta_{\text{s}}\rho_{\text{d}}+3(3w-1)
\Big(6w^3+21w^2+6w-1\Big)\beta_{\text{s}}\rho_{\text{d}}^2\\
&+3(3w-1)\Big(9w^5+69w^4+132w^3+16w^2-29w-5\Big)\text{sign}(\eta)\beta_{\text{s}}\rho_{\text{d}}^3+\gamma \lambda\Big(27w^3+123w^2+97w+17\Big)\beta_{\text{s}}\rho_{\text{d}}^4\\
&+6\lambda^2(9w^4-10w^2+1)\text{sign}(\eta)\beta_{\text{s}}\rho_{\text{d}}^5\Big]\Bigg)+\frac{12\text{sign}(\eta)}{Z}H_{\text{d}}^2\Bigg(1-4\gamma \beta_{\text{s}}\rho_{\text{d}}^2+4(3w-1)^2\Big(1+(w-1)^2\rho_{\text{d}}^2\Big)\beta_{\text{s}}^2\rho_{\text{d}}^2\\
&-4(3w-1)\Big(1-2\gamma\beta_{\text{s}}\rho_{\text{d}}^2\Big)\text{sign}(\eta)\beta_{\text{s}}\rho_{\text{d}}+\frac{6(w+1)^2}{Y^3}(3w-1)\Big(1+(3w+1)\text{sign}(\eta)\rho_{\text{d}}\Big)\Big(1+\big(15w^3+55w^2\\
&-15w-7\big)\rho_{\text{d}}^2+2\Big[5w-1+2(w-1)(3w+1)\lambda \rho_{\text{d}}^2\Big]\text{sign}(\eta)\rho_{\text{d}}\Big)\Big(2(3w-1)\beta_{\text{s}}\rho_{\text{d}}-(1-2\gamma \beta_{\text{s}}\rho_{\text{d}}^2)\text{sign}(\eta)\Big)\beta_{\text{s}}\rho_{\text{d}}\\
&+\frac{4(w+1)}{\text{sign}(\eta) Y}(3w-1)\Big(1+2(w-1)(3w+1)\rho_{\text{d}}^2+(5w-1)\text{sign}(\eta)\rho_{\text{d}}\Big)\Big(1-2\gamma \beta_{\text{s}}\rho_{\text{d}}^2-2(3w-1)\text{sign}(\eta)\times\\
&\beta_{\text{s}}\rho_{\text{d}}\Big)\beta_{\text{s}}\rho_{\text{d}}-\frac{9(3w-1)^2}{Y^2}(w+1)^2\Big((5w-1)\rho_{\text{d}}+\big[1+2(w-1)(3w+1)\rho_{\text{d}}^2\big]\text{sign}(\eta)\Big)^2\beta_{\text{s}}^2\rho_{\text{d}}^2\Bigg)+4w \rho_{\text{d}}+2\Big(1\\
&-\beta_{\text{s}}+3w\big(2\beta_{\text{s}}+w(1-3\beta_{\text{s}})\big)-\big(3w^2-4w+1\big)^2\beta_{\text{s}}\rho_{\text{d}}^2\Big)\text{sign}(\eta)\rho_{\text{d}}^2-4\gamma (3w-1)\beta_{\text{s}}\rho_{\text{d}}^3=0,
\end{split}
\end{equation}
\end{widetext}
Where $Z$ is defined as
\begin{equation}
Z=2(1-3w)\beta_{\text{s}}\rho_{\text{d}}+\big(1-2\gamma\beta_{\text{s}}\rho_{\text{d}}^2\big)\text{sign}(\eta).
\end{equation}
Furthermore, the prime sign indicates differentiation with respect to $t_{\text{d}}$ and $H'_{\text{d}}=\frac{\mid \eta \mid}{\kappa^2} \dot{H}$.  It should be mentioned that even for this  simple toy model, we deal with modified Friedmann equations which are much more complicated compared with the standard Friedmann equations.

Finally, for the dimensionless conservation relation, we have
\begin{equation}\label{rhod}
\rho'_{\text{d}}=-\frac{3}{Y}H_{\text{d}}(w+1)\Big(1+(3w+1)\text{sign}(\eta)\rho_{\text{d}}\Big)\rho_{\text{d}},
\end{equation}
where $\rho'_{\text{d}}=\frac{\mid \eta\mid ^{3/2}}{\kappa^2}\dot{\rho}_{\text{d}}$. In the case of dust ($\omega=0$), Eq. \eqref{rhod} coincides with the standard case, namely $\rho'_{\text{d}}=-3H_{\text{d}}\rho_{\text{d}}$. On the other hand, for an arbitrary equation of state, the continuity equation is singular at $\rho_{\text{d}}^*=-\text{sign}(\eta)/\lambda$. To avoid this singularity we expect that $-\text{sign}(\eta)/\lambda	\notin (0,\rho_{\text{Bd}})$. For example if $-\text{sign}(\eta)/\lambda<0$ then there will not be any singularity in the continuity equation and in the terms proportional to $Y^{-n}$ ($n>0$) in other governing equations. It is easy to verify that for $\omega>-0.13$, only $\eta>0$ is acceptable. Regarding that in the early universe one can assume the dominant components in the cosmic fluid possess a positive equation of sate parameter, then the free parameter $\eta$ should be positive. However, for the sake completeness we explain the $\eta<0$ case alongside with $\eta>0$. Moreover, we cannot certainly rule out the $\eta<0$ case since depending on the magnitude of other free parameters $\rho_{\text{d}}^*$ can be larger than the density at the bounce. Clearly, if this happens, then the continuity equation will be free of singularity.

Eq. \eqref{rhod} can be simply integrated. Let us write its solution for the viable case $\eta>0$
\begin{equation}
a=a_{\min}\Big(\frac{\rho_{\text{Bd}}}{\rho_{\text{d}}}\Big)^{\frac{1}{3 \omega+3}} \Big(\frac{(1+3 \omega) \rho_{\text{Bd}}+1}{(1+3 \omega) \rho_{\text{d}}+1}\Big)^{\frac{\omega (3 \omega+5)}{9 \omega^2+12 \omega+3}}
\end{equation}
where $a_{\min}$ is the scale factor at the bounce. One may plug this equation into Eq. \eqref{Hd} in order to find $a(t)$. However, Eq. \eqref{Hd} cannot be integrated analytically.

As seen from the above relations, in this model, the dynamics of  the cosmic fluid dramatically depends on the magnitude of $\beta_{\text{s}}$ and the choice of the equation of state parameter $w$. Therefore, it is necessary to estimate the order of magnitude of $\beta_{\text{s}}$ and $w$. Furthermore, the sign of the coupling constant $\eta$ is another variable that must be specified. This means that $\beta$ and $\eta$ do not appear separately and only their ratio matters.

Here, we will first attempt to find a constraint on the magnitude of $\beta_{\text{s}}$ and then categorize the possible bouncing solutions in terms of the sign of the coupling constants. In order to constraint $\beta_{\text{s}}$, we need  to estimate $\eta$ and $\beta$. As already mentioned, EMSG should recover the standard gravity at the low energy regime. We use this fact to find a restrict the parameters. In the presence of the cosmological constant, our toy model takes the form $f(R, T)=R+\beta R^2+\eta T^2 +2\Lambda$. Naturally one may expect that the correction terms should be negligible compared to other terms at the present time. This means $\lvert\beta\lvert R^2\ll \Lambda$ and $\lvert\eta\lvert T^2\ll \Lambda$. For the former, by considering that $R\propto \Lambda$, we arrive at $\lvert\beta\lvert\ll {1}/{\Lambda}$. And for the latter, by assuming that $T\simeq\rho_{\Lambda}$ and $\rho_{\Lambda}=\Lambda/\kappa$, we deduce that $\vert\eta\vert\ll {\kappa^2}/{\Lambda}$. So, by gathering these results together, one can show that $\lvert\beta_{\text{s}}\lvert\gg 1$. 

\begin{figure}
\centering
\includegraphics[scale=0.52]{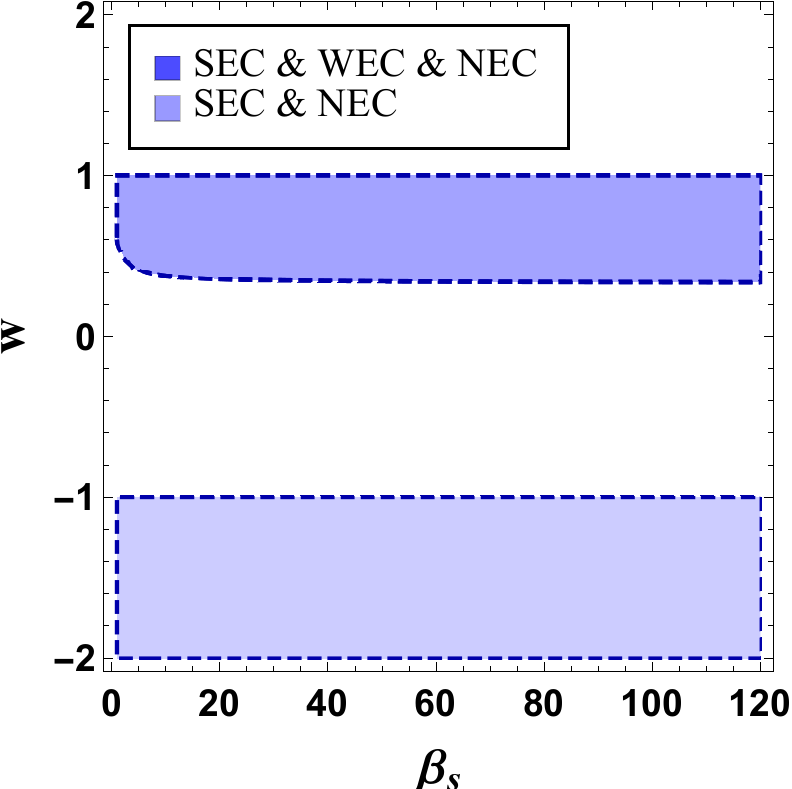}\hspace*{.03cm}
\includegraphics[scale=0.52]{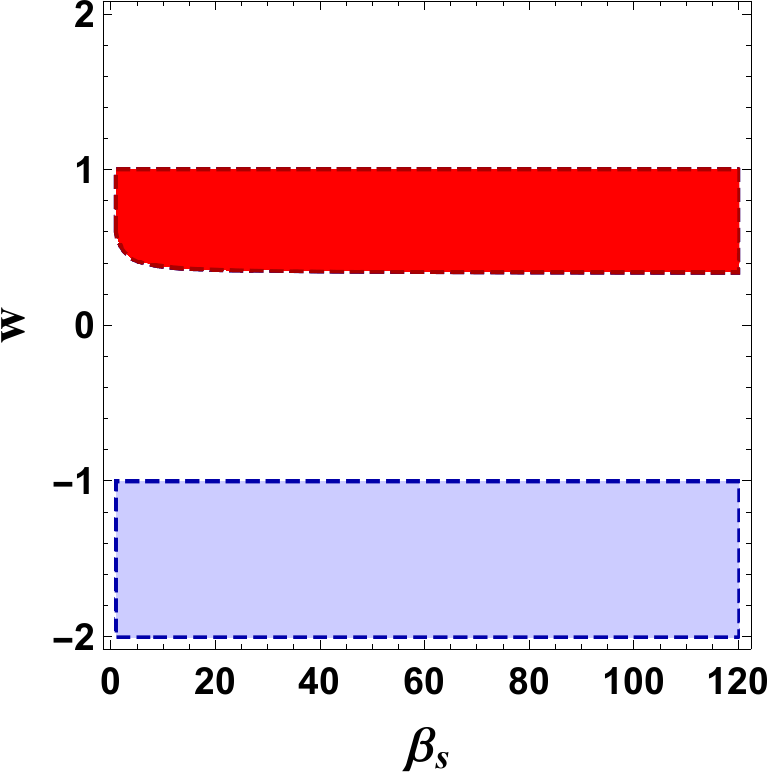}
\vspace*{.03cm}
\includegraphics[scale=0.51]{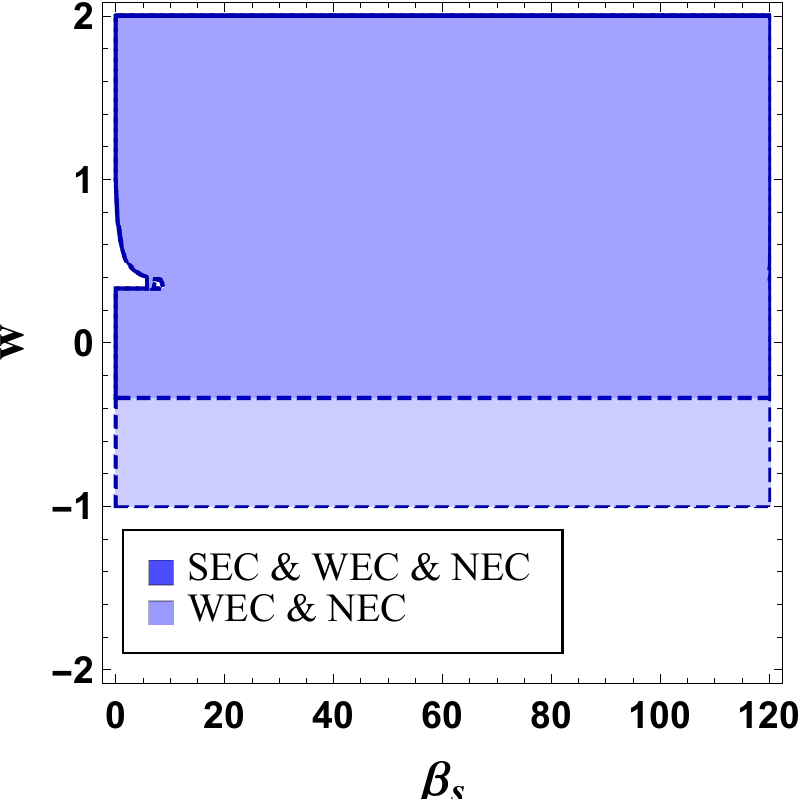}\hspace*{.03cm}
\includegraphics[scale=0.51]{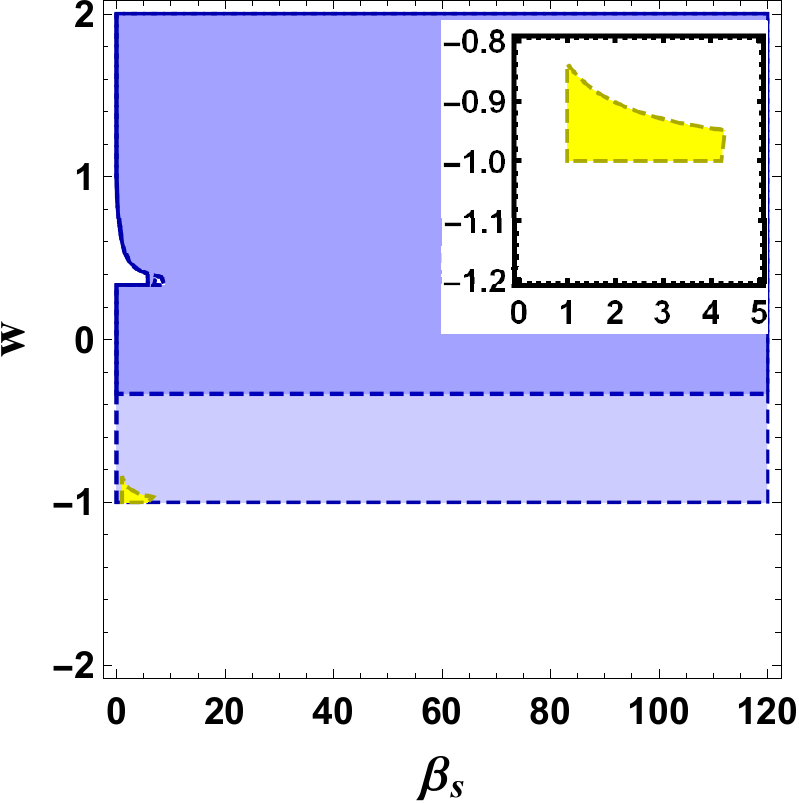}
\caption{The allowed areas for the scaled coupling constant and the equation of state parameter in the $\beta_{\text{s}}-w$ plane for the case $(\eta>0, \beta>0)$. In the top and bottom right panels, the red and yellow zones represent the acceptable values of $\beta_{\text{s}}$ and $w$ for having a cosmic bounce in the Palatini formalism of EMSG. 
These red and yellow areas are obtained from the first and second roots of the $f_{R}$ function, Eqs. \eqref{rho1} and \eqref{rho2}, respectively.
In the top and bottom left panels, for Eqs. \eqref{rho1} and \eqref{rho2}, we respectively exhibit the allowed region where the energy conditions are established.
Here, we assume that $-2<w<2$ and $\beta_{\text{s}}$ is a positive parameter.}\label{Fig1}
\end{figure}

So if $\beta_{\text{s}}$ is a positive quantity, i.e., when $\eta>0$ and $\beta>0$ or $\eta<0$ and $\beta<0$, then $\beta_{\text{s}}\gg 1$. And if $\beta_{\text{s}}$  be a negative quantity, i.e., when $\eta>0$ and $\beta<0$ or $\eta<0$ and $\beta>0$, its reliable realm is $\beta_{\text{s}}\ll -1$. In the following, we classify the bouncing solutions with respect to the sign of $\beta_{\text{s}}$ and in each case illustrate how the sign of the coupling constant $\eta$ as well as the choice of equation of state, i.e., $w$, can affect the existence of a non-singular universe.

As mentioned in the last part of the previous subsection, it is interesting to examine the energy conditions in Palatini EMSG. In the following, for each bouncing solution, we check these conditions.

\subsubsection{Case I: If $\boldsymbol{\beta_{\text{s}}>0}$}
\begin{center}
IA: $\eta>0$ and $\beta>0$
\end{center}

We first focus on the case where $\eta>0$ and $\beta>0$.
Back in Subsec. \ref{General bouncing conditions}, we found the required conditions for the existence of bounce.  In a similar way, our first task here is to find the roots of the dimensionless Friedmann equation \eqref{Hd} in which we put $\text{sign}(\eta)=+1$. 
By considering A-C cases in the previous subsection, we see that the two cases $f_{R}=0$ and $\Delta=0$ can induce cosmic bounces. Therefore, in this model, we restrict our study to the $f_{R}$-induced and $\Delta$-induced bounces. 
\begin{figure}
\centering
\includegraphics[scale=0.65]{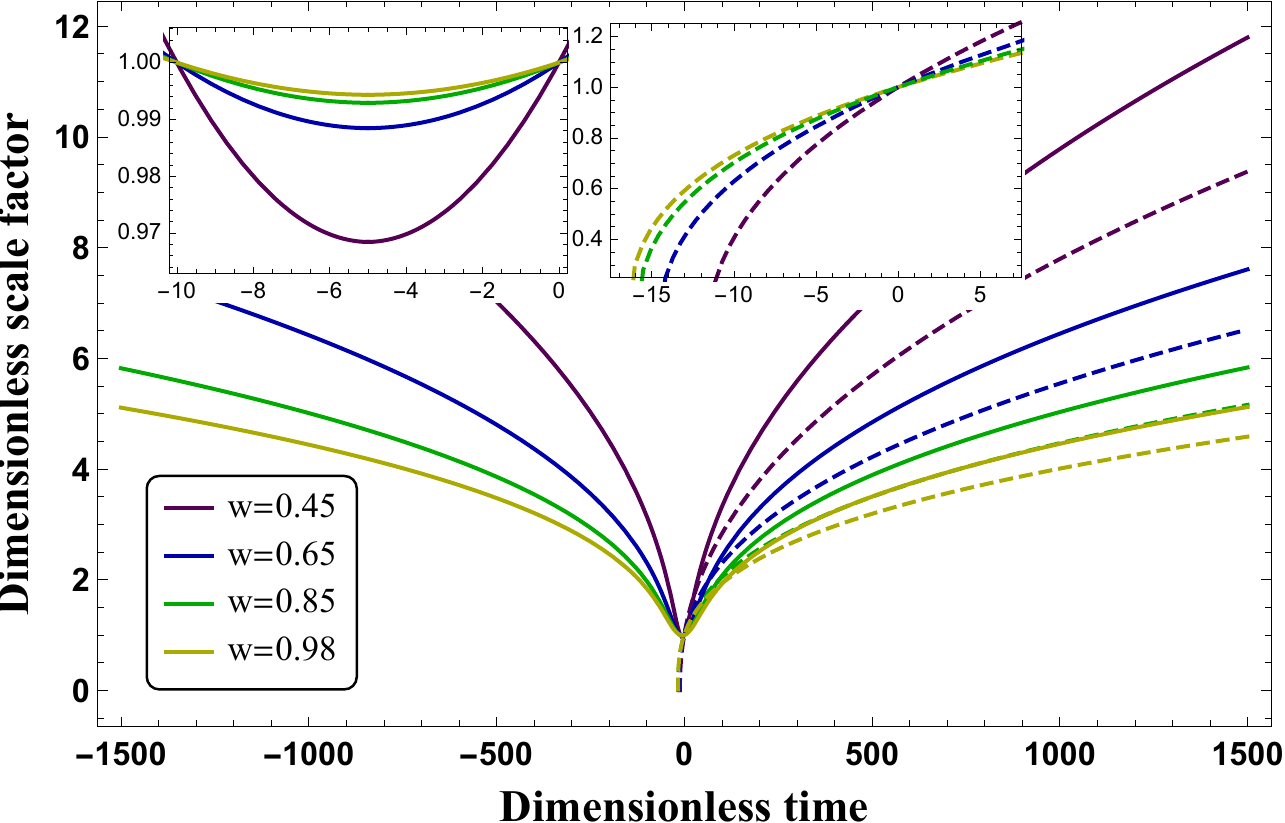}\vspace{1cm}
\includegraphics[scale=0.65]{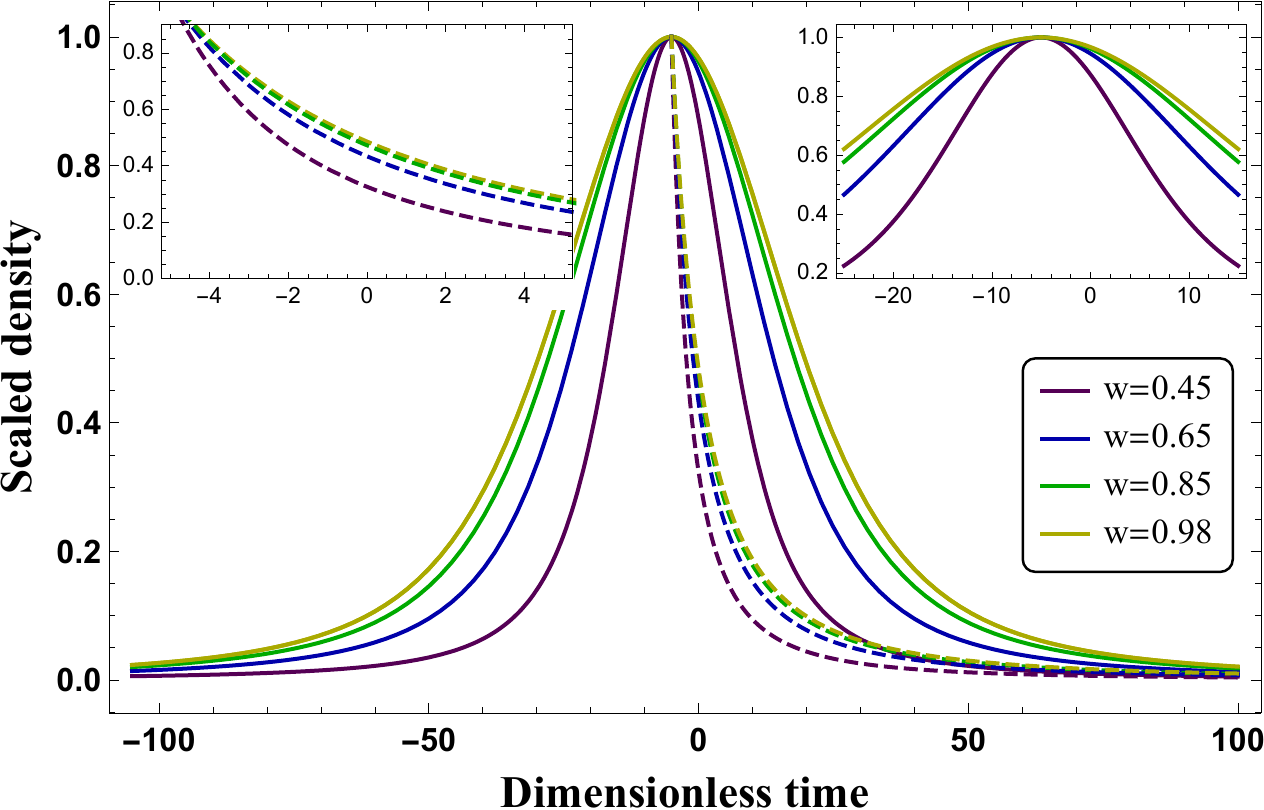}\vspace{1cm}
\includegraphics[scale=0.65]{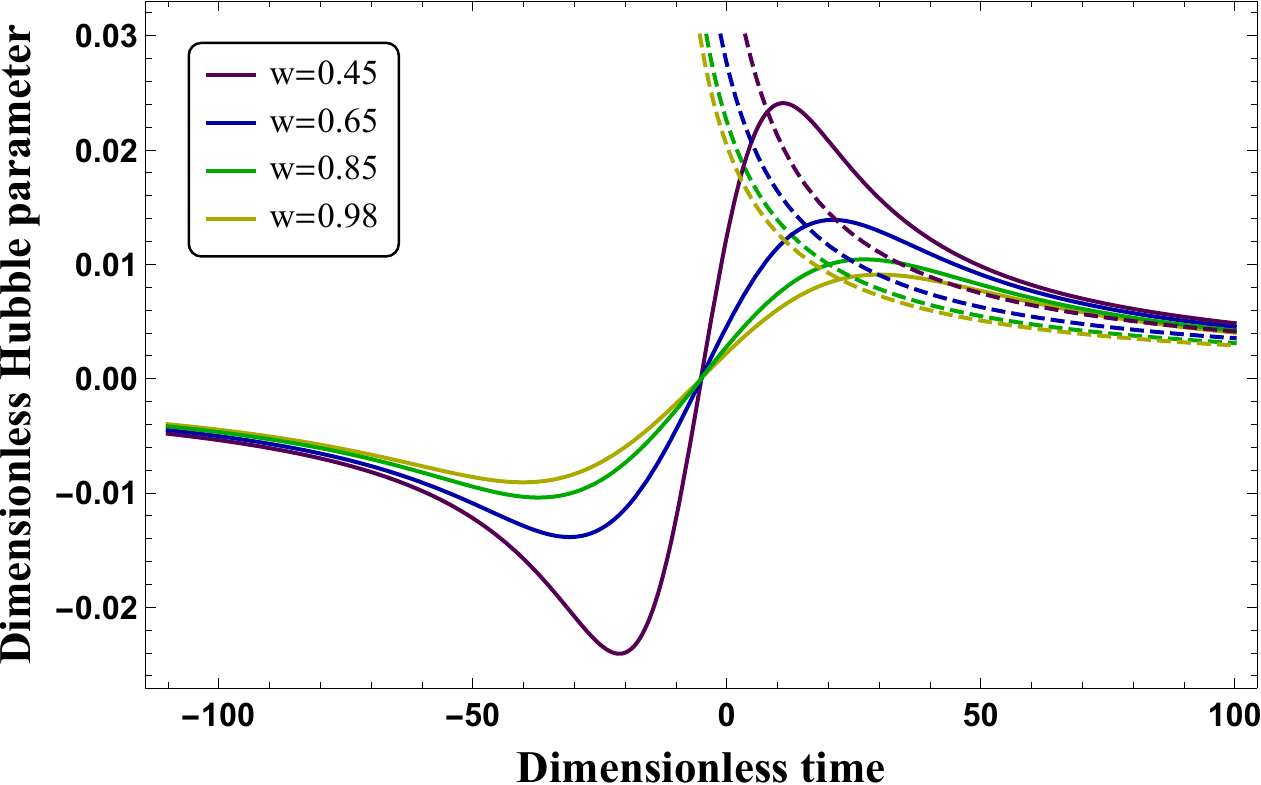}\vspace{1cm}
\caption{The dynamics of the perfect cosmic fluid in the very early universe in Palatini EMSG in which we choose that $\eta>0$ and $\beta>0$. Here, we assume that $\beta_{\text{s}}=100$. From top to bottom, the panels respectively illustrate the evolution of the dimensionless scale factor, scaled energy density, and dimensionless Hubble parameter in terms of $t_{\text{d}}$ for different values of $w$ selected from the red region in Fig.  \ref{Fig1}. Here, for $w=0.45, 0.65, 0.85, 0.98$, we obtain $\rho_{Bd}=0.0144, 0.0053, 0.0032, 0.0026$, respectively. In each case, the analytical GR solutions, dashed curves, are also scaled over the corresponding $\rho_{Bd}$.}\label{Fig2}
\end{figure}
For the $f_{R}$ and $\Delta$ functions, we respectively obtain two and three roots that can describe a bounce. Then, by inserting each root within Eqs. \eqref{cond-2} and \eqref{cond-3}, and imposing the second to fifth bouncing conditions introduced in Subsec. \ref{General bouncing conditions}, we obtain allowed region in the $\beta_{\text{s}}-w$ plane.
Here, we assume that $\beta_{\text{s}}> 1$ and $-2<w<2$. For the current case, these allowed areas are exhibited in Fig. \ref{Fig1}.
In the top and bottom right panels, the yellow and red surfaces represent the allowed areas that come from two roots of the $f_{R}$ function. Considering $\beta_{\text{s}}\gg 1$ reveals that the yellow region is not physically important since $\beta_s$ is not large enough. In this case, for the roots of $\Delta$ function, we do not have an allowed regime in this interval of $\beta_{\text{s}}$ and $w$. 
It should be noted that to display these areas, we also impose another condition under which each root should be a real and positive parameter to describe a physical energy density. 
In this case, the roots of the $f_{R}$ function are given by 
\begin{subequations}
\begin{align}
\label{rho1}
&\rho_{\text{d}1}=\frac{1}{(3w-1)\beta_{\text{s}}-\sqrt{A}},\\
\label{rho2}
&\rho_{\text{d}2}=\frac{1}{(3w-1)\beta_{\text{s}}+\sqrt{A}},
\end{align}
\end{subequations}
where $A=(3w-1)\big(w(2+3\beta_{\text{s}})-\beta_{\text{s}}-2\big)\beta_{\text{s}}$.
We should also mention that the roots of the $\Delta$ function are long and for some values of $\beta_{\text{s}}$ and $w$, they can be imaginary. So, let us only provide one of them as an example
\begin{align}
&\rho_{\text{d}3}=-\frac{1}{3B}\Big(3+\frac{2B}{w-1}+w^2(\beta_{\text{s}}-1)\\
\nonumber
&~~~~~~~+\frac{B^2}{(1-4w+3w^2)^2\beta_{\text{s}}}+\beta_{\text{s}}-6w(4+\beta_{\text{s}})\Big).
\end{align}
Here
\begin{align}
\nonumber
&B=\Big[3\sqrt{3}C+(1-3w)^4(w-1)^3\beta_{\text{s}}^2\Big(36+9w^2(3+\beta_{\text{s}})\\
&~~~+3w\big(15-2\beta_{\text{s}}\big)+\beta_{\text{s}}\Big)\Big]^{1/3},
\end{align}
in which
\begin{align}
& C=(1-3w)^3(w-1)^3\beta_{\text{s}}^{3/2}\Big(\big(1+w(8+3w)\big)^3\\\nonumber
&+(1-3w)^2\big(18w^4+42w^3+77w^2+104w+47\big)\beta_{\text{s}}\\\nonumber
&+3\big(1-3w\big)^4\big(1+w\big)^2\beta_{\text{s}}^2\Big)^{1/2}.
\end{align}

Now, by regarding these solutions and considering Eqs. \eqref{NEC}-\eqref{SEC}, we are in a position to examine the energy conditions. In the top and bottom left panels of Fig. \ref{Fig1}, for Eqs. \eqref{rho1} and \eqref{rho2}, we respectively exhibit the allowed region in the $\beta_{\text{s}}-w$ plane where the energy conditions are established. As can be seen from the comparison of the top row panels in this figure, for the first root of $f_{\text{R}}$ function, i.e., Eq. \eqref{rho1}, the allowed region for having a cosmic bounce is completely restricted to the area where all NEC, WEC, and SEC are established. On the other hand, for the second root of $f_{\text{R}}$ function, i.e., Eq. \eqref{rho2}, the corresponding area is a very tiny part of the zone where only NEC and WEC are satisfied. See the yellow surface. Therefore, in the latter case, SEC is violated.

So far, the acceptable area for the scaled coupling constant $\beta_{\text{s}}$ and the equation of state parameter $w$ have been obtained. Now, by choosing a suitable point $\big(\beta_{\text{s}}, w\big)$\footnote{ This point should not be on the border of the allowed area.} from this allowed area and inserting that within the definition of $H_{\text{d}}^2$, we obtain a maximum energy density that can truly describe a universe without the Big Bang singularity. We display this dimensionless density by $\rho_{\text{Bd}}$.
For instance, by choosing the allowed point $\big(92, 0.8\big)$ from the red area of the top right panel of Fig. \ref{Fig1} and inserting it within Eq. \eqref{Hd}, we obtain the numerical value of the roots of $H_{\text{d}}$. Among these roots, only the smallest positive real root, here $\rho_{\text{d}}=0.004$, satisfies all five bouncing conditions \textit{(i)-(v)} introduced in Subsec. \ref{General bouncing conditions}. We call this value the density at the bounce.

Therefore, for this case, $\rho_{\text{Bd}}=0.004$. One should note that the allowed regions in the $\beta_{\text{s}}-w$ plane satisfy all the bouncing conditions and after choosing a suitable point from this plane, as illustrated above, one can find the density at the bounce.

\begin{figure}
\centering
\includegraphics[scale=0.67]{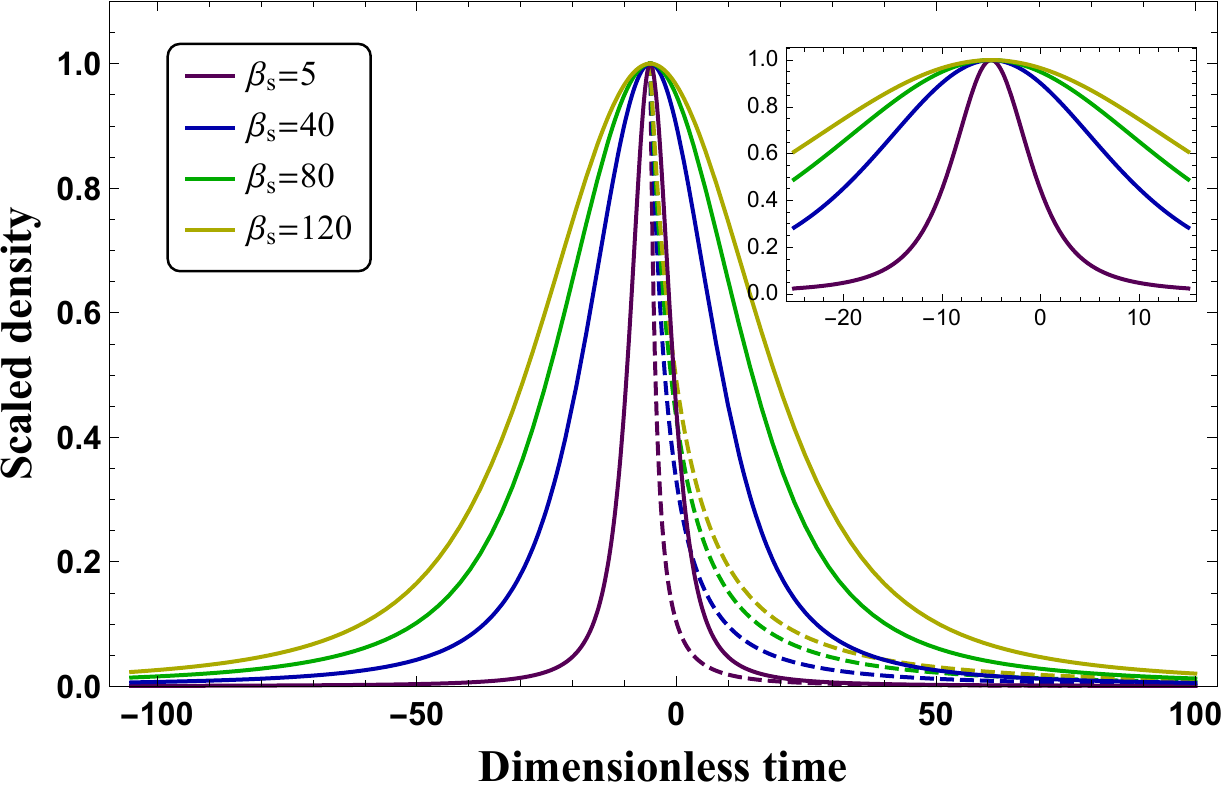}\vspace{1cm}
\includegraphics[scale=0.7]{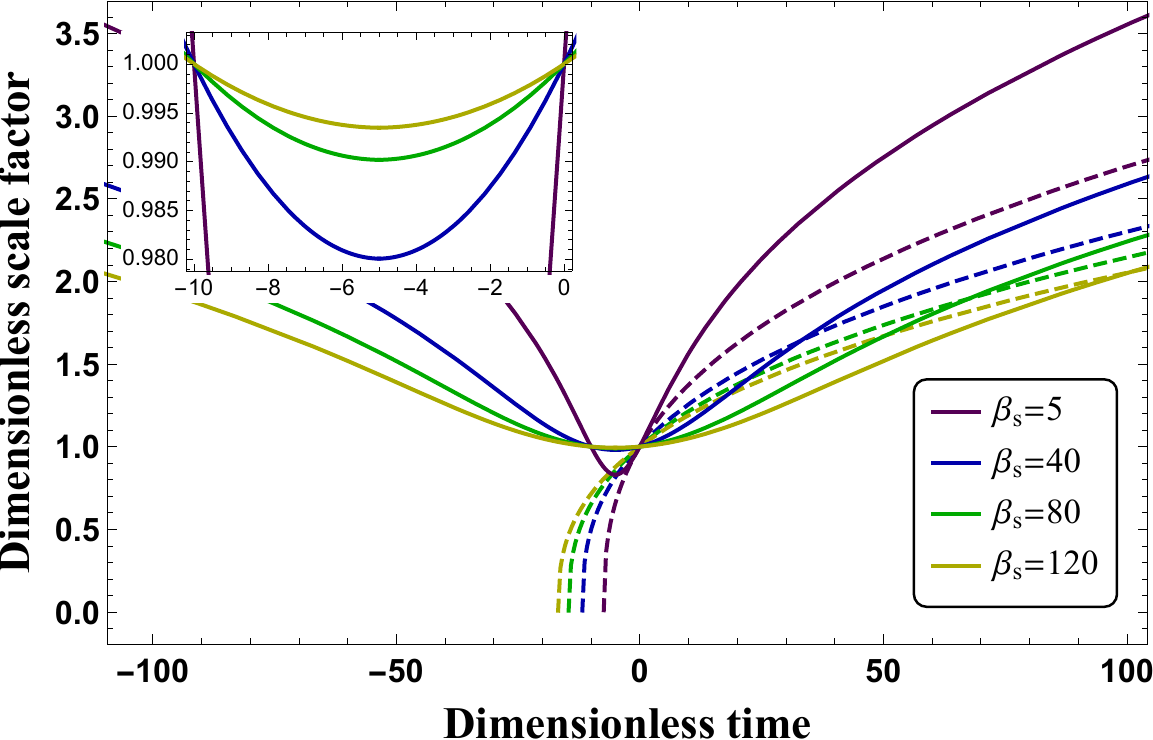}\vspace{1cm}
\includegraphics[scale=0.68]{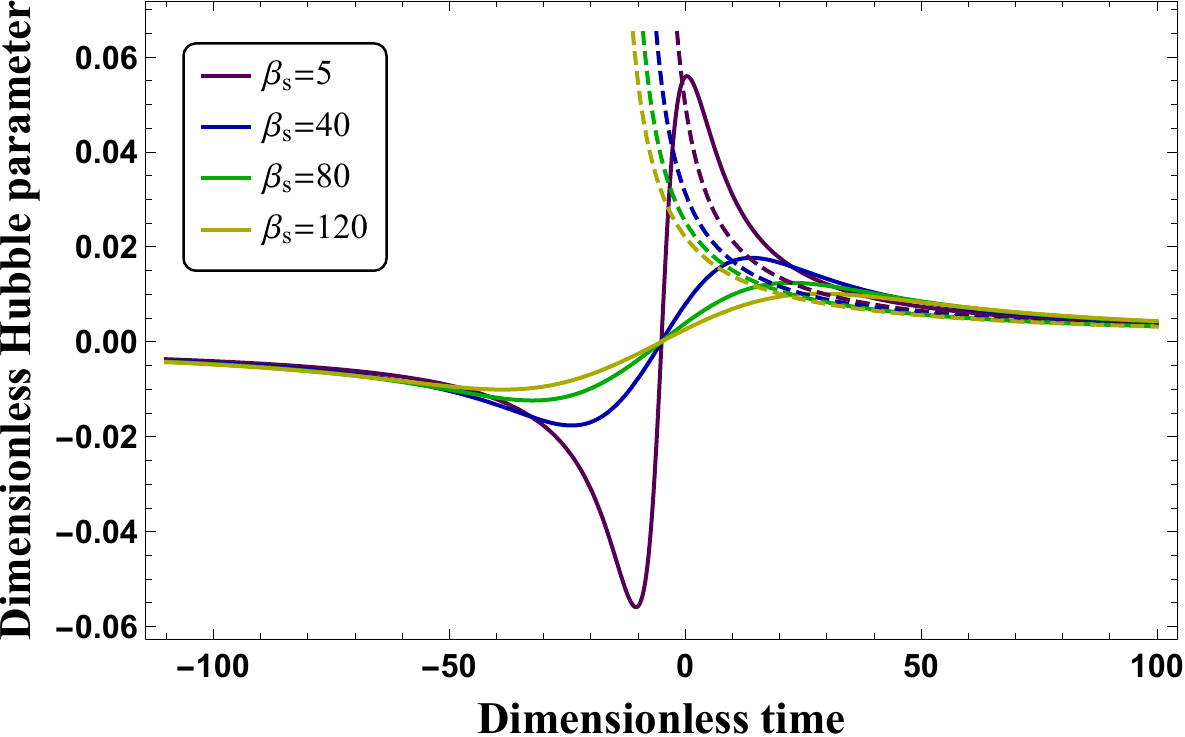}\vspace{1cm}
\caption{The dynamics of the perfect cosmic fluid in the very early universe in Palatini EMSG in which we consider that $\eta>0$ and $\beta>0$. Here, we assume that $w=0.8$. From top to bottom, the panels illustrate the behavior of the scaled energy density, dimensionless scale factor, and dimensionless Hubble parameter in terms of $t_{\text{d}}$ for different values of $\beta_{\text{s}}$, respectively.
Here, for $\beta_{\text{s}}=5, 40, 80, 120 $, we obtain $\rho_{Bd}=0.072, 0.009, 0.005, 0.003$, respectively. In each case, the analytical GR solutions, dashed curves, are also scaled over the corresponding $\rho_{Bd}$.}\label{Fig3} 
\end{figure}

After rescaling $\rho_{\text{d}}$ by $\rho_{\text{Bd}}$, we numerically solve Eqs. \eqref{Hd} and \eqref{rhod} for the allowed corresponding values of $\beta_{\text{s}}$ and $w$, and find the dimensionless energy density and scale factor in terms of dimensionless time.
We repeat this method for any point selected from the allowed area. 
These results are exhibited in Figs. \ref{Fig2} and \ref{Fig3}. Here, for convenience, we display the time evolution of $\rho_{\text{s}}$. Therefore, for all cases at the bounce we have $\rho_s=1$.

In  Figs. \ref{Fig2} and \ref{Fig3}, the solid and dashed lines express the dynamics of the cosmic fluid in Palatini EMSG and GR, respectively. In these figures, one can see that there is a cosmic bounce for each allowed value of $\beta_{\text{s}}$ and $w$ selected from the red region of the top right panel in Fig.  \ref{Fig1}. Here, it is assumed that the bounce occurs at $t_{\text{Bd}}=-5$.
In fact, without lose of generality, we set $t_{\text{d}}=-5$ as the bouncing time and leave $t_{\text{d}}=0$ as the initial time in this scenario. Of course from the physical point of view, the time origin is not important.

We see that at the bounce the scale factor undergoes a minimum. At this minimum, the scale factor is nonzero. On the other hand, as expected the energy density has a finite value maximum at the bounce. It should be mentioned that for all of these plots, the same initial and physical conditions are considered. We assume that $a_{\text{d}}(t_{\text{d}}=0)=1$ and $\rho_{\text{s}}(t_{\text{d}}=t_{\text{Bd}})=1$. Also, in order to achieve a meaningful comparison, the GR plots are derived from the same conditions and in each case, they are scaled by the corresponding value of $\rho_{\text{Bd}}$. In the following, we will explain more about this issue.

For a detailed survey of the dynamics of $a_{\text{d}}$ and $\rho_{\text{s}}$, in Figs. \ref{Fig2} and \ref{Fig3}, we study their evolution in terms of different magnitudes for $w$ and $\beta_{\text{s}}$. Here, the density varies smoothly in the interval $\rho_{\text{s}}\in (0, 1)$. This guarantees that before and after the bounce there is no singularity in the governing equations, i.e., the third and fifth bouncing conditions introduced in Subsec. \ref{General bouncing conditions} are truly established. So these solutions can be viable bounces.

As seen from the top panel of Fig. \ref{Fig2}, by decreasing $w$, the dimensionless scale factor evolves faster, i.e., at a fixed time, the slope of a tangent line to these curves increases with decreasing $w$.
It should be noted that in this case, $w=1/3$ does not lie in the allowed region, see the red region of the top right panel of Fig. \ref{Fig1}. In the next case, where $w$ can be very close to $w=1/3$, we will discuss this specific value of $w$ which can clearly indicate the radiation dominated phase of the thermal history of the universe.

It should be noted that the dashed curves (in these figures and forthcoming ones) are analytically derived in the GR context.
In this standard case, the Friedmann equations have the exact solution. Let us first write these equations in the dimensionless form. The result is
\begin{equation}
\dot{\rho}_{\text{s}}+3(1+w)H_\text{d} \rho_{\text{s}}=0, ~~~~~H_\text{d} =\sqrt{\frac{\rho_{\text{Bd}}\rho_{\text{s}}}{3}}.
\end{equation}
Regarding the above mentioned initial conditions, the solution takes the following form
\begin{subequations}
\begin{align}
a(t_{\text{d}})=\bigg(\frac{1}{5}t_{\text{d}}\Big(1-\sqrt{\rho_0}\Big)+1\bigg)^{\frac{2}{3(1+w)}},
\end{align}
\end{subequations}
and
\begin{subequations}
\begin{align}
\rho_{\text{s}}(t_{\text{d}})=\frac{25 \rho_0}{\Big(t_{\text{d}}\big(\sqrt{\rho_0}-1\big)-5\Big)^2}, 
\end{align}
\end{subequations}
where $\rho_0=\rho_{\text{s}}(t_{\text{d}}=0)$ and is given by
\begin{subequations}
\begin{align}
\rho_0= \bigg(\frac{2}{2+5\sqrt{3}\sqrt{\rho_{\text{Bd}}}(1+w)}\bigg)^2.
\end{align}
\end{subequations}

Considering these results in the vicinity of the bounce, see the magnified figure at the top right of the upper panel of Fig. \ref{Fig2}, one may deduce that the GR cosmic fluid with larger $w$s, and also smaller $\rho_{\text{Bd}}$s, has a larger dimensionless scale factor at a fixed time. Furthermore, by decreasing $w$, the GR cosmic fluid expands more rapidly in the vicinity of the bounce. This fact can be also grasped from this magnified figure. 
In the middle panel of Fig. \ref{Fig2}, we study the time evolution of $\rho_{\text{s}}$ in terms of different values of $w$. As seen, in EMSG, the cosmological fluid with smaller $w$ expands much faster than those with larger $w$. The GR system behaves in the same manner.
This fact is illustrated in the zoomed figure at the top left of this panel. We recall that the above studies are only reliable in the vicinity of the bounce.

Moreover, to complete our study, we survey the behavior of the dimensionless Hubble parameter for each case. As it is obvious, apart from the vicinity of the bounce, like the GR case, $H_{\text{d}}$ is a decreasing function in terms of the time. This fact is also investigated for the next solutions.

The behavior of $\rho_{\text{s}}$, $a_{\text{d}}$, and $H_{\text{d}}$ in terms of different $\beta_{\text{s}}$ is also examined in Fig. \ref{Fig3}. Here, we choose a fixed $w$ and several $\beta_{\text{s}}$s from the allowed area in the top right panel of Fig. \ref{Fig1}. One can see that by decreasing $\beta_{\text{s}}$, the cosmic energy density decreases more quickly. Furthermore, at the smaller $\beta_{\text{s}}$, the universe expands faster.

We should keep it in mind if the bounces exist in EMSG, they do not necessarily occur at the GR singularity, and they might just be in its vicinity. This fact is clarified in Figs. \ref{Fig2} and \ref{Fig3}. 
As one can see from these figures, the GR singularity does not occur exactly at $t_{\text{Bd}}=-5$. This fact is also checked for the bouncing solutions studied in the next part. However, one should note that the location of the bounce depends on our choice for the initial conditions. In other words, it is always possible to choose initial conditions in such a way that GR singularity and EMSG bounce happens at the same $t_{\text{d}}$.

\begin{center}
IB: $\eta<0$ and $\beta<0$
\end{center}

We next turn to study the other possible case, i.e., when both $\beta$ and $\eta$ are negative. For this case, by considering $\text{sign}(\eta)=-1$ in Eqs. \eqref{Hd}-\eqref{rhod}, and applying the same method as before, the allowed values of $\beta_{\text{s}}$ and $w$ can be obtained. In this case, all roots of the $f_{R}$ function as well as two of the roots of the $\Delta$ function do not specify any area in the $\beta_{\text{s}}-w$ plane. Moreover, 
for one of the roots of the $\Delta$ function, we obtain several very small regions most of which are only displayed with their borders. As the chosen point from these areas should not be on the border, we discard these unphysical regions. In fact, for very specific choices of the parameters from these tiny areas, one may obtain a bounce. But this seems like a fine-tuning and is not satisfactory. Therefore, we ignore these kinds of unreliable solutions.
To sum up, in this case where $\beta<0$ and $\eta<0$, there is no allowed value of $\big(\beta_{\text{s}}, w\big)$ for which the regular cosmic bounce can exist. However, we do not rule out the possibility of the existence of a viable cosmic bounce in the case with $\beta<0$ until we check the other case $\beta<0$ and $\eta>0$ in the next part.

\subsubsection{Case II: If $\boldsymbol{\beta_{\text{s}}<0}$}

Here, we discuss the conditions for existing a cosmic bounce in the case $\beta_{\text{s}}<0$. For this case, we have shown that $\beta_{\text{s}}$ should satisfy the physical range $\beta_{\text{s}}\ll -1$. Here, as before, we choose that $-2<w<2$. By considering the definition of $\beta_{\text{s}}$, two cases $(\eta<0, \beta>0)$ and $(\eta>0, \beta<0)$ are possible. So, in the following, we investigate them separately and study the existence of the bounce in each one. For both cases, again by considering the sign of $\eta$, we find the roots of Eq. \eqref{Hd}. Then, for each root, we impose the second to fifth bouncing conditions introduced in Subsec. \ref{General bouncing conditions}. Furthermore, as before, we consider that $\rho_{\text{root}}$ should be real and positive. By regarding these conditions, we find the allowed region for $w$ and $\beta_{\text{s}}$ that may discard the Big Bang singularity and lead to a bounce in the very early universe. Next, by applying these values, we obtain the numerical value of the smallest real positive root of $H_{\text{d}}$ as $\rho_{\text{Bd}}$.
Finally, we numerically study the dynamics of $a_{\text{d}}$ and $\rho_{\text{s}}$ for the corresponding allowed values of $w$ and $\beta_{\text{s}}$. Here, we consider the physical and initial conditions similar to the preceding part. 

\begin{figure}
\centering
\includegraphics[scale=0.53]{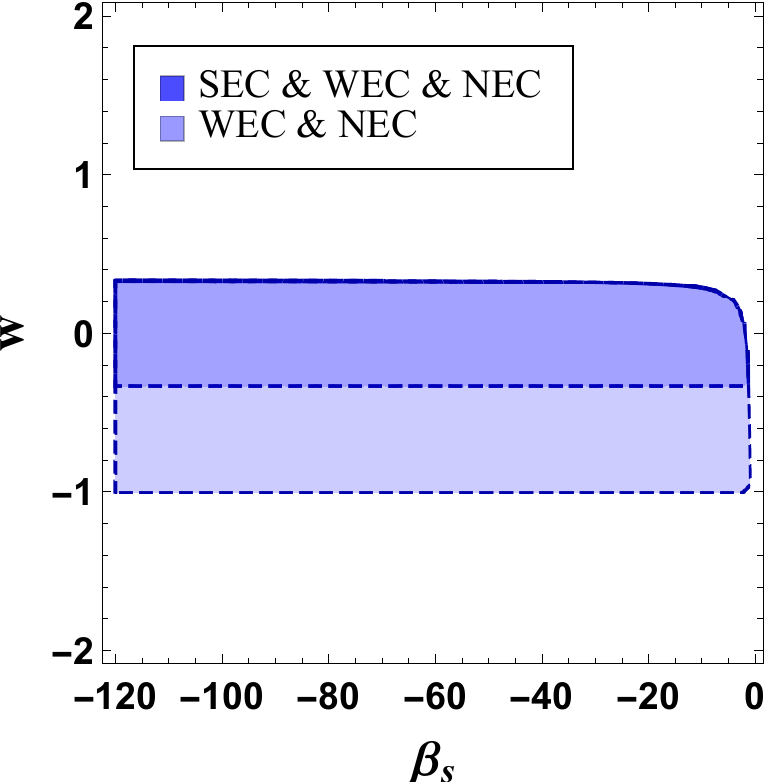}\hspace*{.03cm}
\includegraphics[scale=0.55]{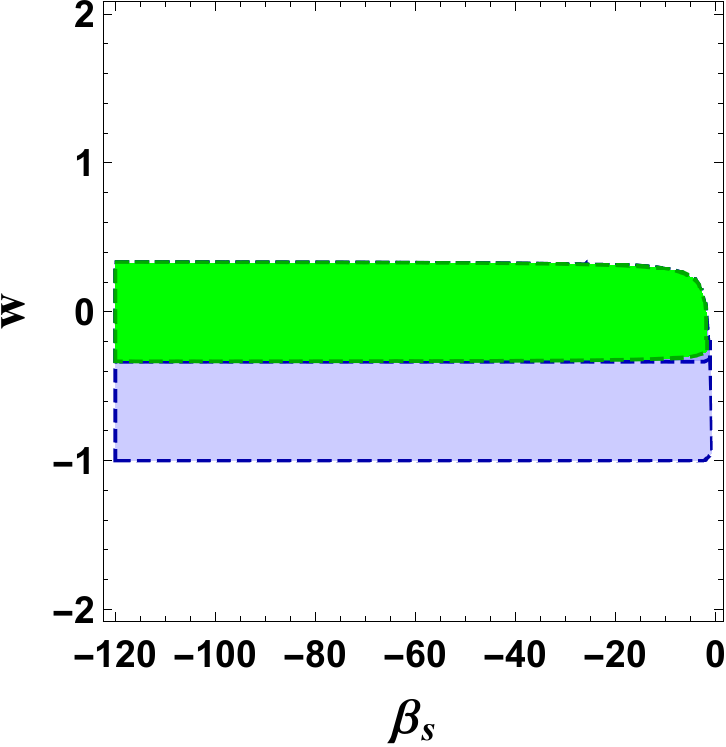}
\caption{The allowed area for the scaled coupling constant and the equation of state parameter in the $\beta_{\text{s}}-w$ plane. 
In the right panel, the green surface represent the acceptable values of $\beta_{\text{s}}$ and $w$ for having a cosmic bounce in the Palatini formalism of EMSG. 
The this panel belongs to the case $(\eta>0, \beta<0)$. 
In this case, the green surface is deduced from one root of the $f_{R}$ function.
In the left panel, for the corresponding roots, we exhibit the allowed region where the energy conditions \eqref{NEC}-\eqref{SEC} are established.
Here, we assume that $-2<w<2$ and $\beta_{\text{s}}$ is a negative parameter.}\label{Fig5}
\end{figure}
\begin{center}
IIA: $\eta<0$ and $\beta>0$
\end{center}
For the case of $\eta<0$ and $\beta>0$, by regarding the above method, one can derive the allowed areas of $\beta_{\text{s}}$ and $w$.
In this case, the $f_R=0$ condition does not specify any area in the $\beta_{\text{s}}-w$ plane.
Also, the regions obtained from the $\Delta=0$ condition are quite tiny. Therefore, as before, we drop these artificial areas. So, in this case, similar to the case IB where both $\eta$ and $\beta$ are negative, there is no allowed surface in the $\beta_{\text{s}}-w$ plane that can describe the non-singular universe. In fact, we found that although there is a bounce for which most of the required conditions are satisfied. However, at least one of the governing equations gets singular somewhere in the range ($0,\rho_{\text{Bd}}$).
\begin{figure}
\centering
\includegraphics[scale=0.7]{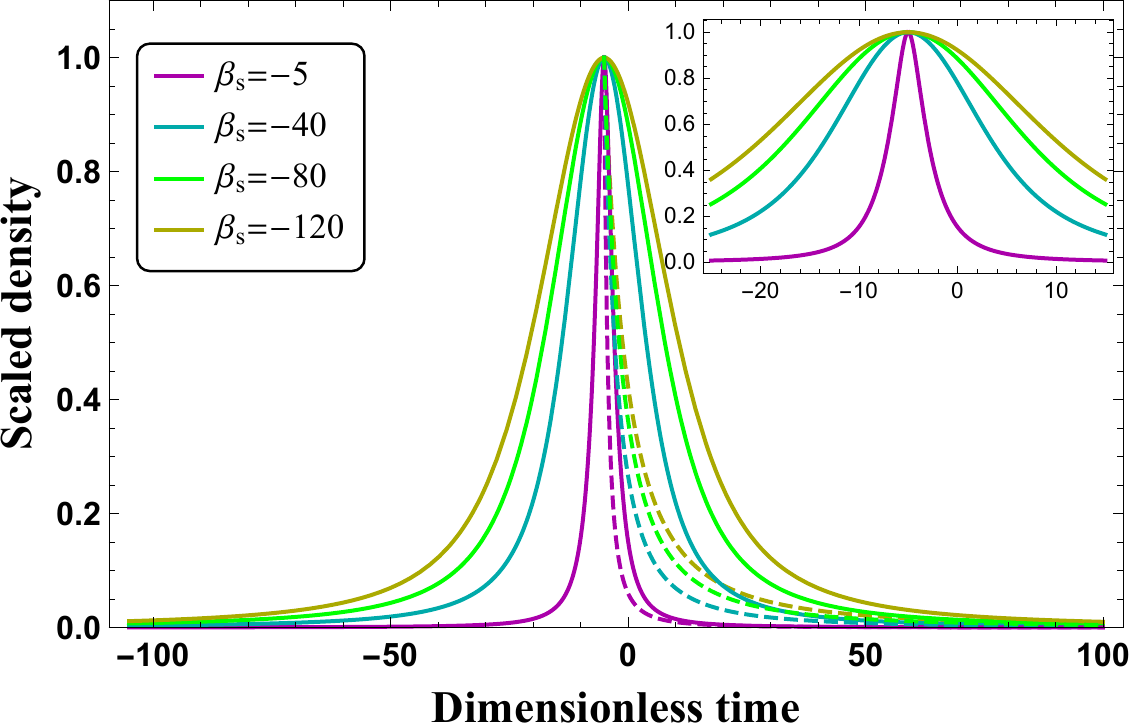}\vspace{1cm}
\includegraphics[scale=0.7]{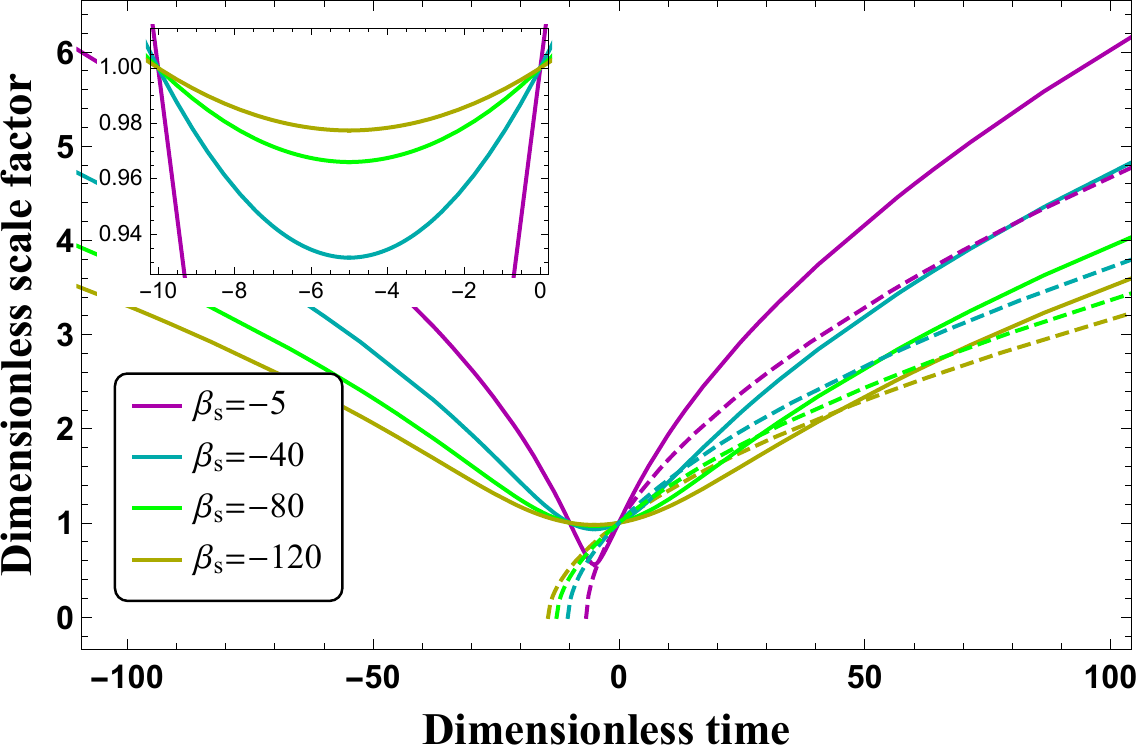}\vspace{1cm}
\includegraphics[scale=0.7]{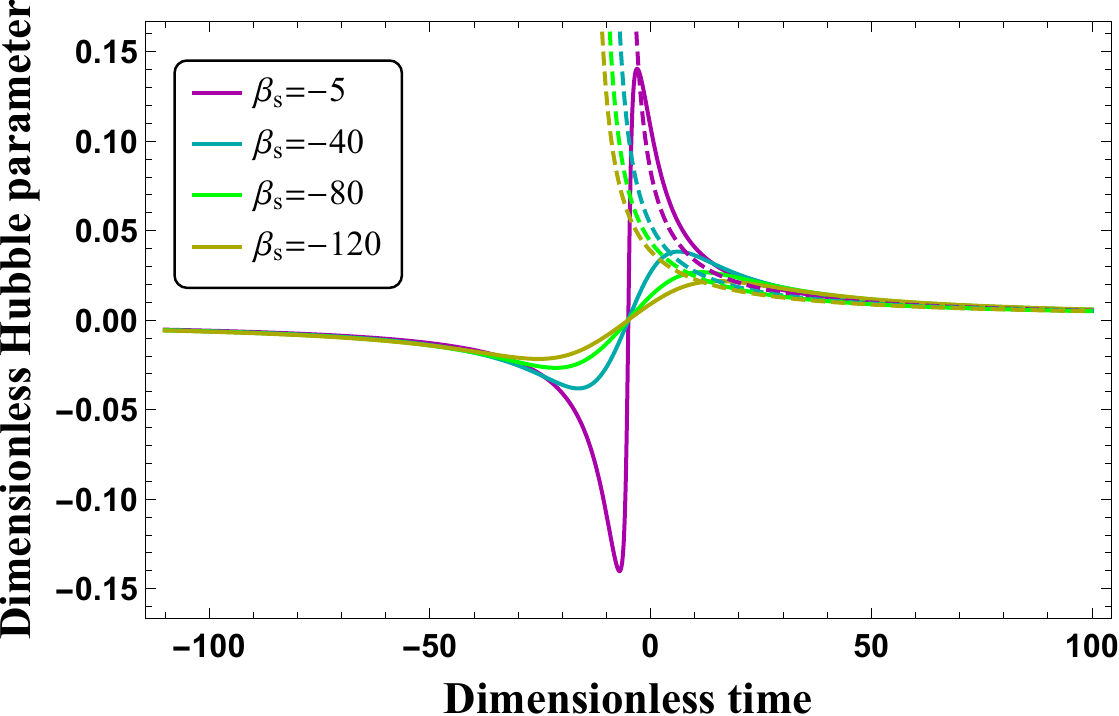}\vspace{1cm}
\caption{The dynamics of the perfect cosmic fluid in the very early universe in Palatini EMSG in which we consider that $\eta>0$ and $\beta<0$. Here, we assume that $w=0.2$.
From top to bottom, the panels illustrate the behavior of the scaled energy density, dimensionless scale factor, and dimensionless Hubble parameter in terms of $t_{\text{d}}$ for different values of $\beta_{\text{s}}$, respectively.
These values are selected from the green area in the bottom right panel of Fig. \ref{Fig5}. Here, for $\beta_{\text{s}}=-5, -40, -80, -120 $, we obtain $\rho_{Bd}=0.35, 0.03, 0.02, 0.01$, respectively. In each case, the analytical GR solutions, dashed curves, are also scaled over the corresponding $\rho_{Bd}$.
}\label{Fig7}
\end{figure}

\begin{center}
IIB:  $\eta>0$ and $\beta<0$
\end{center}

\begin{figure}
\centering
\includegraphics[scale=0.66]{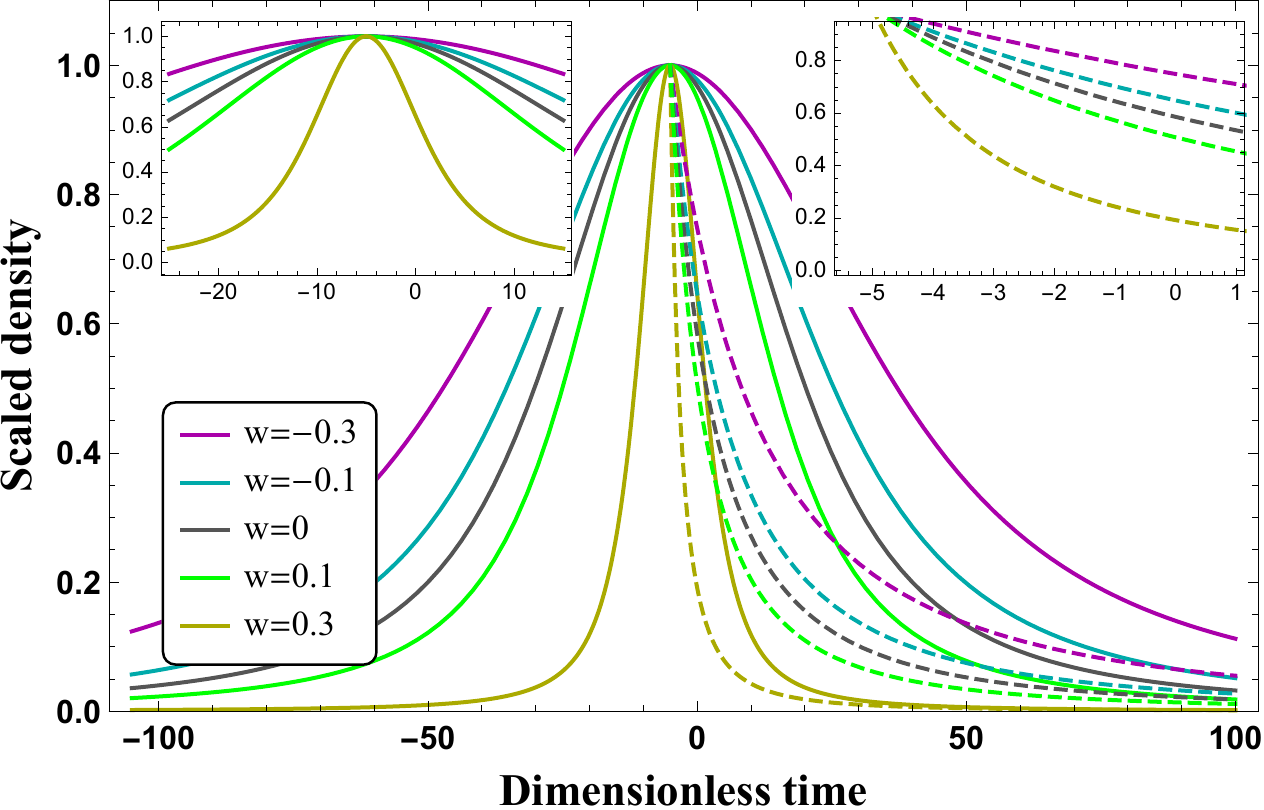}\vspace{1cm}
\includegraphics[scale=0.68]{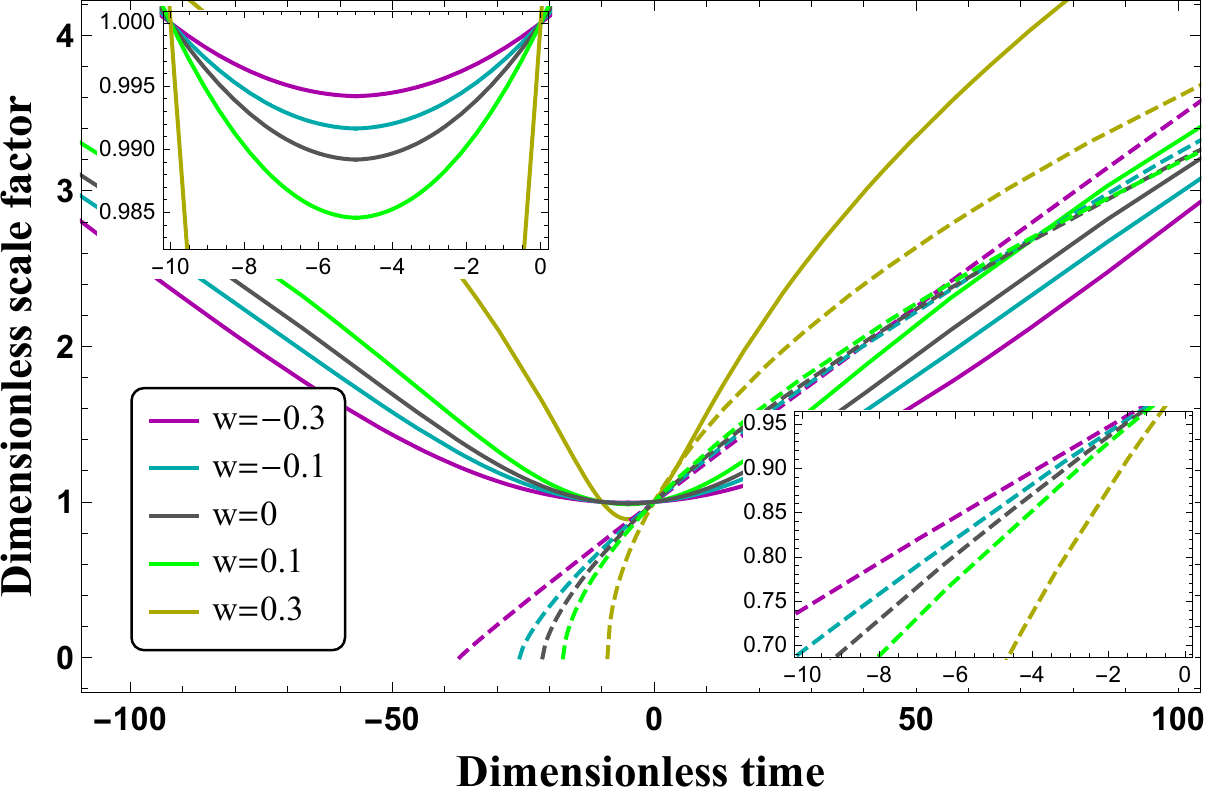}\vspace{1cm}
\includegraphics[scale=0.66]{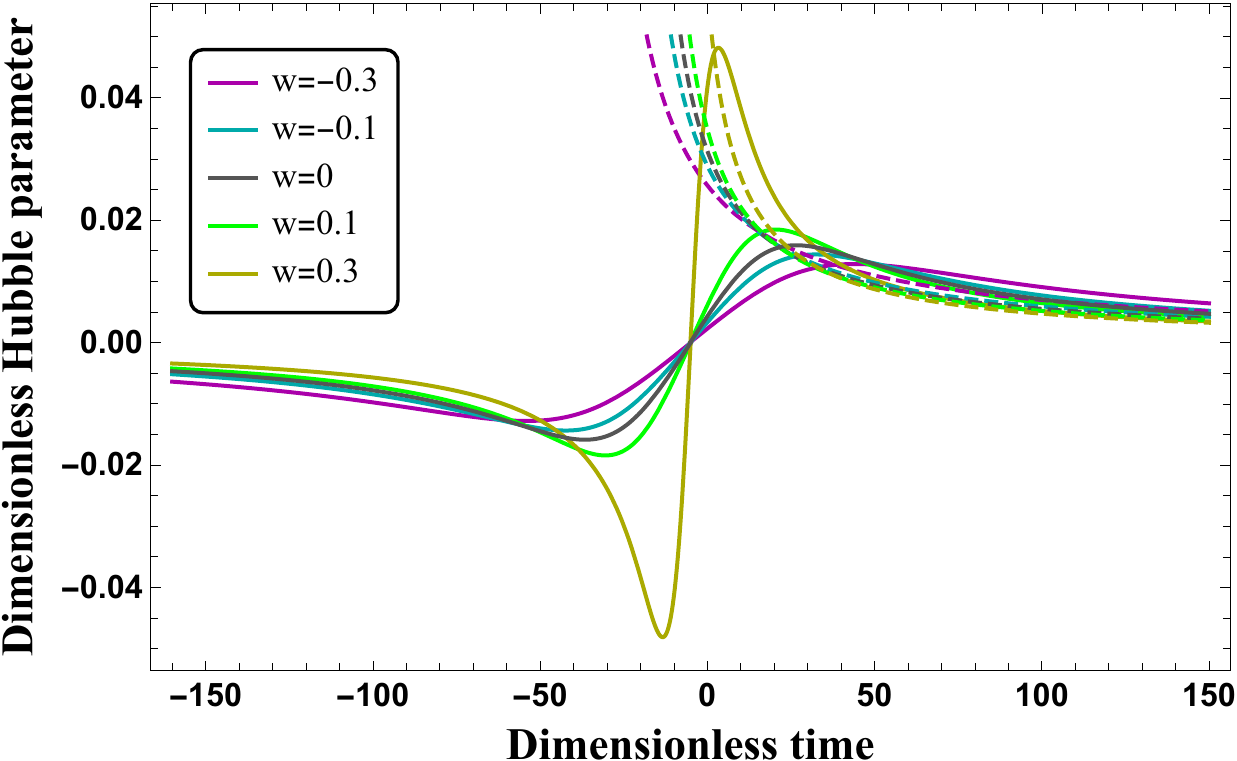}\vspace{1cm}
\caption{ The dynamics of the perfect cosmic fluid in the very early universe in Palatini EMSG in which we choose that $\eta>0$ and $\beta<0$. Here, we assume that $\beta_{\text{s}}=-100$. From top to bottom, the panels illustrate the evolution of the scaled energy density, dimensionless scale factor, and dimensionless Hubble parameter in terms of $t_{\text{d}}$ for different values of $w$, respectively. 
These values are selected from the green area in the bottom right panel of Fig. \ref{Fig5}. Here, for $w=-0.3, -0.1, 0, 0.1, 0.3$, we obtain $\rho_{Bd}=0.003, 0.004, 0.005, 0.007, 0.052$, respectively. In each case, the analytical GR solutions, dashed curves, are also scaled over the corresponding $\rho_{Bd}$.}\label{Fig8}
\end{figure}

In the next probable case where $\eta>0$ and $\beta<0$, the allowed zones are shown in Fig. \ref{Fig5} and the bouncing solutions are illustrated in Figs. \ref{Fig7} and \ref{Fig8}. First of all, let us mention that the smooth behavior of the density guarantees that there is no singularity in this case.
Here, the green surface is deduced from one of the roots of the $f_{R}$ function.  For the rest of the roots of $f_{R}$ and also $\Delta$ functions, we do not achieve an allowed regime in this interval of $\beta_{\text{s}}$ and $w$.
In Fig. \ref{Fig7}, by fixing the equation of state, we examine the behavior of $\rho_{\text{s}}$, $a_{\text{d}}$, and $H_{\text{d}}$ in terms of $\beta_{\text{s}}$, which is negative here. It is obvious that for larger $\beta_{\text{s}}$, the energy density of the cosmic fluid vanishes faster, and the infant universe grows more rapidly. 
This behavior contrasts with the $\beta_{\text{s}}>0$ case.
In Fig. \ref{Fig8}, by fixing the value of $\beta_{\text{s}}$, we also see that by increasing $w$, the cosmological fluid decreases more quickly and it expands with higher speed.
The same behavior is observed in GR in the vicinity of the bounce point.
These facts can be easily grasped from the magnified plots in the top and middle panels of Fig. \ref{Fig8}.
This behavior is also quite different from the previous case where $\beta_{\text{s}}>0$. We will discuss these differences in more detail in the next subsection.

\subsection{Discussion}

Now, let us summarize all the main results obtained in the preceding subsections. First of all, as seen, the dimensionless energy density starts from infinity in GR, while, for some specific cases in the Palatini formalism of EMSG, this function has a finite maximum, namely $\rho_{\text{B}}$. On the other hand, the cosmic scale factor undergoes a nonzero minimum. This means that Palatini EMSG provides bouncing solutions. In all the viable cases, we saw that the density varies more slowly near the bounce compared with the corresponding model in GR.

The second point is that the sub-case $\eta<0$ in the both cases $\beta_{\text{s}}>0$ and $\beta_{\text{s}}<0$ provides no region in the parameter space $\beta_{\text{s}}-w$ for the existence of the bounce. 
From this perspective, one may conclude that only the sub-cases $(\eta>0,\beta<0)$ and $(\eta>0, \beta>0)$ are physically interesting.

The next is that, by considering the definition of the scaled coupling constant and the results obtained from both Figs. \ref{Fig3} and \ref{Fig7}, after assuming a fixed $\beta$, one can deduce that near the bounce, a cosmic fluid with larger $\eta$ expands faster than that with smaller $\eta$. Accordingly, the density falls more rapidly.

Before discussing the fourth point, let us recall the definition of fluid stiffness. As we know, the stiffness of the fluid is measured by the sound speed $c_{\text{s}}$, e.g., see \cite{rezzolla2013relativistic}.
For simplicity, here, we consider the definition of the sound speed in the Newtonian regime, i.e., $c_{\text{s}}^2=d p/d \rho$ where we assume that the pressure is only a function of $\rho$.
By regarding the barotropic equation of state $p=w\rho$, one may deduce that $w$ can represent the stiffness of the fluid. Therefore, the big value of $w$ indicating the dramatic changes in pressure for small changes in density refers to a stiff equation of state and the small value of $w$ refers to a soft equation of state.

Now, by considering the above definition, we interpret the results of the previous subsection. As seen from Fig. \ref{Fig2}, in both GR and Palatini contexts, near the Big Bang singularity, softer cosmic fluids expands more quickly than stiffer ones. This behavior is observed in the case $\beta_{\text{s}}>0$.
So if the cosmic fluid has soft and stiff components and they were around the bouncing point, the softer component would expand faster.  
On the other hand, this behavior is completely different in the case $\beta_{\text{s}}<0$. In this case, as seen from Fig. \ref{Fig8}, soft components are more resistant to expansion than those with a stiff equation of state.

The last remark is to emphasize the existence of the bouncing cosmology for interesting toy models. As one can see from the numerical curves in Fig. \ref{Fig8}, a cosmic bounce can even exist for the cold dust with $w=0$ and the fluid with $w\simeq 1/3$. We reiterate again that the exact $\omega=1/3$ case does not lie in the allowed region for the existence of bounce in our toy model. However, the equation of state for the bouncing solution can be very close to $\omega=1/3$. In this figure, it is also shown that a cosmic fluid with a negative equation of state that may be a model for the dark energy, can have a bouncing solution.
Furthermore, for these toy models,  it is illustrated that the energy density of the radiation-dominated universe with $w\simeq 1/3$ falls faster than that of the matter-dominated universe. And the latter one disappears more quickly than the energy density of the universe with a negative $w$. So if the component of the cosmic fluid with a negative pressure could exist in the vicinity of the bounce, it would be more dominant than the rest of the components of the cosmic fluid in the late-time universe. Of course, these results are true only for this simple toy model, and for more realistic results, more physical models should be considered.

\section{Summary and Conclusion} \label{sec:con}

In this paper, we have studied the Palatini formulation of EMSG. In this formalism, the modified version of the field equations has been derived. We have explored their consequences in different contexts. We have shown that the energy-momentum tensor $T_{\mu\nu}$ is not conserved and consequently the path of a test particle deviates from the standard geodesic curves of the spacetime. In fact, in this context, a test particle undergoes an extra force called the fifth force. We showed that in the vacuum, the geodesic equations coincide with the standard case. This means that EMSG does not predict any fifth force in the vacuum.

We have also investigated the weak field limit of this formalism and obtained the modified version of the Poisson equation. It is expected that the corrections in the modified Poisson equation of EMSG are negligible even in the galactic scales. As already mentioned, the original version of EMSG is neither a theory for dark matter nor for dark energy. It is worth mentioning that in the weak field limit of this theory, the left-hand side of the field equations is reduced to the formal wave equation. See Eq. \eqref{star}. Therefore, by investigating the right-hand side of this wave equation and completing the analysis of the linearized theory in Palatini EMSG, one can study the gravitational waves in this theory \cite{NR}.

In the second part of this paper, we have explored the cosmological behavior of this theory. In fact, in order to obtain the deviation of EMSG from GR, we focused on the evolution of the cosmos at very early times. To do so, we have obtained the modified version of the Friedmann equations in Palatini EMSG.
We have next introduced the general bouncing conditions and then by choosing a toy model as $f(R, Q)=R+\beta R^2+\eta Q$, as the simplest Palatini EMSG model, we have examined the bouncing solutions. To ensure that at low curvature regime, this model recovers GR, the coefficient of the linear term $R$ is unity and the coupling constants $\beta$ and $\eta$ satisfy the constraint $|\eta|\ll\kappa^2/\Lambda$ and $|\beta|\ll 1/\Lambda$.
Like the metric formulation of EMSG, here, it has been shown that the cosmic fluid has a finite energy density value in the very early universe and crosses a bounce. On the other hand, the cosmic scale factor has a nonzero minimum. Our study on the bouncing solution is in favor of the case $\eta>0$. On the other hand, as the case $\eta<0$ does not provide any allowed area, we rule out the possibility of the existence of a regular bounce for the cases where $\eta<0$.
We have also shown that most of the acceptable values of $\beta_{\text{s}}$ and $w$ for having a cosmic bounce in the Palatini formalism of EMSG, satisfy all three energy conditions WEC, NEC, and SEC. It has been exhibited that near the bounce, a cosmic fluid with larger $\eta$ expands faster than that with smaller $\eta$. Accordingly, the density falls more rapidly. Furthermore, in all cases, the density varies more slowly near the bounce compared with the corresponding model in GR.

Therefore, in Palatini EMSG, the regular bounce exists and the cosmic fluid dramatically deviates from the GR case at the Big Bang singularity. This fact leads us to a considerable question: Does this formalism possess the true sequence of the cosmological epochs? As it is shown in \cite{roshan2016energy}, a powerful method to check this issue is to use the dynamical system analysis. In future work, we plan to study the dynamical system approach to the cosmological evolution in Palatini EMSG. Another crucial study that is necessary to be done in EMSG, is the stability of the model against tensorial cosmological perturbations. It is known that bouncing cosmological models, like EiBI, suffer from tensor instabilities \cite{EscamillaRivera:2012vz}. If this kind of instability happens in EMSG, then a serious challenge will be raised for the viability of the model. We leave this issue as a subject for future studies.

\section*{acknowledgments}
We are grateful to {\"O}zg{\"u}r Akarsu for constructive comments. Also we appreciate Fatimah Shojai for very fruitful discussions and suggestions that substantially improved the early version of this paper. This work is supported by Ferdowsi University of Mashhad.


\end{document}